\documentclass[12pt]{article}
\usepackage{tocloft}
\usepackage{amsfonts}
\usepackage{amsmath}
\usepackage{amssymb}
\usepackage{array,slashed}
\usepackage{bigints}
\usepackage{bm}
\usepackage{bbm}
\usepackage{upgreek}
\usepackage{booktabs}
\usepackage[nosort]{cite}
\usepackage{color}
\usepackage{dsfont}
\usepackage{float}
\usepackage{framed}
\usepackage{graphicx}
\usepackage{indentfirst}
\usepackage{mathrsfs}
\usepackage{multirow}
\usepackage{pdflscape}
\usepackage{setspace}
\usepackage{subdepth}
\usepackage{subfig}
\usepackage{titlesec}
\usepackage{wrapfig}
\usepackage[all]{xy}
\usepackage{young}
\usepackage[vcentermath]{youngtab}
\usepackage{relsize}
\usepackage{stackengine}
\usepackage{verbatim}
\usepackage{slashed}

\newcommand{\re}{{\rm e}}
\newcommand{\ri}{{\rm i}}
\newcommand{\rd}{{\rm d}}

\usepackage{hyperref}
\hypersetup{colorlinks=true}
\hypersetup{linkcolor=black}
\hypersetup{citecolor=black}
\hypersetup{urlcolor=black}

\numberwithin{equation}{section}

\usepackage[left=2.5cm,right=2.5cm,top=2.5cm,bottom=3cm]{geometry}
\fontdimen2\font=1.2\fontdimen2{\jot}{5pt}

\titleformat{\section}{\bfseries}{\thesection.}{4pt}{}
\titlespacing{\section}{0pt}{20pt}{6pt}

\titleformat{\subsection}{\normalfont\itshape}{\thesubsection.}{4pt}{}
\titlespacing{\subsection}{0pt}{15pt}{6pt}

\titleformat{\subsubsection}{\normalfont\itshape}{\thesubsubsection.}{4pt}{}
\titlespacing{\subsubsection}{0pt}{15pt}{6pt}

\titleformat{\paragraph}{\normalfont\itshape}{\theparagraph.}{4pt}{}
\titlespacing{\paragraph}{0pt}{15pt}{6pt}

\def\tilde{\widetilde}

\def\bar{\overline}

\def\half{{1 \over 2}}

\def\1{{\mathds 1}}
\def\Re{\mathop{\rm Re}}
\def\Im{\mathop{\rm Im}}

\DeclareMathOperator{\Tr}{\mathrm{Tr}}
\DeclareMathOperator{\Arg}{Arg}
\DeclareMathAlphabet{\mathbfsf}{OT1}{cmss}{bx}{n}

\newcommand{\C}{{\mathbb C}}
\newcommand{\IC}{{\mathbb C}}

\newcommand{\ba}{\begin{aligned}}
\newcommand{\ea}{\end{aligned}}

\def\bZ{\mathbb{Z}}

\newcommand{\cC}{\mathcal{C}}

\newcommand{\cH}{\mathcal{H}}
\newcommand{\cI}{\mathcal{I}}
\newcommand{\cJ}{\mathcal{J}}
\newcommand{\cK}{\mathcal{K}}

\newcommand{\cN}{\mathcal{N}}
\newcommand{\cO}{\mathcal{O}}

\newcommand{\ts}{\textrm{s}}
\newcommand{\tw}{\textrm{w}}

\def\SU{\mathrm{SU}}

\def\U{\mathrm{U}}

\newcommand{\ed}{\,.}
\newcommand{\ec}{\,,}

\newcommand{\be}{\begin{equation}}
\newcommand{\ee}{\end{equation}}

\newcommand{\AdS}{\textrm{AdS}}

\newcommand{\HP}{\textrm{HP}}
\newcommand{\BPS}{\textrm{BPS}}

\DeclareFontShape{OT1}{cmr}{mx}{n}%
{<->cmr10}{}
\newcommand{\mytitlefont}{\fontseries{mx}\selectfont}
\DeclareMathAlphabet{\titlemath}{OT1}{cmr}{mx}{n}

\begin{document}

%
\begin{titlepage}
\begin{center}
~\\[2cm]
{\fontsize{29pt}{0pt} \mytitlefont Delayed Deconfinement and the Hawking-Page Transition }
~\\[1cm]
Christian Copetti,\,$^{1,2}$ Alba Grassi,\,$^{2,3}$ Zohar Komargodski,\,$^2$ and Luigi Tizzano\,$^2$\hskip1pt
~\\[0.5cm]
$^1$~{\it Instituto de Fisica Teorica UAM/CSIC, c/Nicolas Cabrera 13-15, \\Universidad Autonoma de
Madrid, Cantoblanco, 28049 Madrid, Spain}\\[0.2cm]
			
$^2$\,{\it Simons Center for Geometry and Physics, SUNY, Stony Brook, NY 11794, USA}\\[0.2cm]

$^3$\,{\it Institut f\"{u}r Theoretische Physik, ETH Z\"{u}rich, CH-8093 Z\"{u}rich, Switzerland}\\[0.2cm]
\vskip0.8cm
			
\end{center}
\noindent 

\noindent We revisit the confinement/deconfinement transition in ${\cal N}=4$ super Yang-Mills (SYM) theory and its relation to the Hawking-Page transition in gravity. Recently there has been substantial progress on counting the microstates of 1/16-BPS extremal black holes. However, there is presently a mismatch  between the Hawking-Page transition and its avatar in ${\cal N}=4$ SYM. This led to  speculations about the existence of new gravitational saddles that would resolve the mismatch. Here we exhibit a phenomenon in complex matrix models which we call ``delayed deconfinement.'' 
It turns out that when the action is complex, due to destructive interference, tachyonic modes do not necessarily condense. We demonstrate this phenomenon in ordinary integrals, a simple unitary matrix model, and finally in the context of ${\cal N}=4$ SYM. Delayed deconfinement implies a first-order transition, in contrast to the more familiar cases of higher-order transitions in unitary matrix models. We determine the deconfinement line and find remarkable agreement with the prediction of gravity.
On the way, we derive some results about the Gross-Witten-Wadia model with complex couplings.
Our techniques apply to a wide variety of (SUSY and non-SUSY) gauge theories though in this paper we only discuss the case of ${\cal N}=4$ SYM.

\vfill 
\begin{flushleft}
August 2020
\end{flushleft}
\end{titlepage}
%
		
	
\setcounter{tocdepth}{3}
\renewcommand{\cfttoctitlefont}{\large\bfseries}
\renewcommand{\cftsecaftersnum}{.}
\renewcommand{\cftsubsecaftersnum}{.}
\renewcommand{\cftsubsubsecaftersnum}{.}
\renewcommand{\cftdotsep}{6}
\renewcommand\contentsname{\centerline{Contents}}
	
\tableofcontents


\section{Introduction}
In 3+1 dimensional Conformal Field Theory (CFT) the problem of counting local operators is equivalent to the problem of counting states in the Hilbert space $\mathcal{H}_{S^3}$ of the theory on $S^3$. More elegantly, we may Wick rotate to Euclidean signature, compactify the time direction, and study the partition function on $S^3\times S^1$:
$$Z_{S^3\times S^1} = \sum_{\Delta} e^{-\Delta \beta/R}~,$$
where $R$ is the radius of $S^3$ and $\beta/2\pi$ is the radius of the $S^1$. The sum over $\Delta$ runs over all the scaling dimensions of local operators in the CFT. One can decorate this partition function with chemical potentials for global symmetries. One particularly interesting chemical potential that will be important below is the insertion of $(-1)^F$, which leads to alternating signs in the partition function, depending on whether the corresponding state is bosonic or fermionic.

Of course, in interacting theories this partition sum is too complicated to evaluate. Here we will be interested in theories with $\mathcal{N}=1$ superconformal supersymmetry, where some simplifications occur once we add appropriate chemical potentials. From now on, we set $R=1$ without loss of generality. If there is a supercharge $\mathcal{Q}$ that furnishes a supersymmetry of the theory on $S^3\times \mathbb{R}$ and if  $Q_i$ are some ordinary conserved charges which commute with $\mathcal{Q}$, then the partition function 
$$\mathcal{I}=\Tr_{\cH_{S^3}}  (-1)^F \re^{-\delta \{\mathcal{Q},\mathcal{Q}^\dagger\} } \re^{- i \mu_i Q_i}~,$$ has some interesting properties. Since bosons and fermions cancel for nonzero $\{\mathcal{Q},\mathcal{Q}^\dagger\}$, the partition function is in fact $\delta$-independent and only receives contributions from states annihilated by both $\mathcal{Q}$, and $\mathcal{Q}^\dagger$.

For theories with $\mathcal{N}=1$ supersymmetry, one can choose a supercharge such that $\{\mathcal{Q},\mathcal{Q}^\dagger\}\sim \Delta-2j_1+{3\over 2 }r$, with $j_1$ the angular momentum with respect to the left $SU(2)$ acting on $S^3$ and $r$ is the superconformal $R$-charge. We may call this Hilbert space where $\Delta-2j_1+{3\over 2 }r=0$, $\cH_{\BPS}$. One can think of it as a $1/4$-BPS Hilbert space. There are two additional canonical chemical potentials which can be always added, corresponding to the right-moving angular momentum $j_2$ and a combination of $\Delta$ and $j_1$. One therefore arrives at a function $\mathcal{I}$ of two parameters (for a review along with references to the extensive literature on the subject see~\cite{Rastelli:2016tbz,Gadde:2020yah})
\begin{equation}\label{Index}\mathcal{I} = \Tr_{\cH_{\BPS}} (-1)^F p^{{1\over 3} (\Delta+j_1)+j_2}q^{{1\over 3} (\Delta+j_1)-j_2}~. \end{equation}
This index can be thought of as a partition function on $S^3\times S^1$ with appropriate background gauge fields turned on. Then $p,q$ are interpreted as the two complex structure parameters on the space with topology $S^3\times S^1$~\cite{Closset:2013vra,Closset:2014uda}. Globally, the superconformal index is defined on some cover of the space of complex structures. For $\cN=4$ SYM, as we will see, it turns out to live on a triple cover of the space of complex structure moduli. This is perhaps analogous to the discussion in~\cite{Seiberg:2018ntt}.

If we consider $|p|=|q|$ then this corresponds to a round $S^3$ base. One can then adopt the general parameterization $p=e^{-\beta+i\sigma_1}$, $q=e^{-\beta+i\sigma_2}$ where $\sigma_{1,2}$ are real parameters corresponding to the fibration of $S^1$ over $S^3$. For instance, if $\sigma_1=\sigma_2=0$ then the corresponding manifold $S^3\times S^1$ is now a direct product space with $S^1$ of length $\beta$. The index then takes the form 
\begin{equation}\label{OldCIndex}\mathcal{I} = \Tr_{\cH_{\BPS}} (-1)^F e^{-{\beta\over 3} (2\Delta+2j_1)}=\Tr_{\cH_{\BPS}} (-1)^F e^{-\beta (\Delta+{1\over 2}r)}~.\end{equation}
The limit $\beta\to 0^+$ then simply counts all the BPS operators with signs. This is the limit of small $S^1$ with no fibration over $S^3$. 

The index~\eqref{Index} can be calculated in many conformal gauge theories in four dimensions, which is possible because supersymmetry guarantees that the index is independent of the exactly marginal parameters (and hence can be typically computed at zero coupling). Our primary interest here would be $\mathcal{N}=4$ maximally supersymmetric Yang-Mills theory, which from the point of view of $\mathcal{N}=1$ supersymmetry has an ordinary (non-$R$) global $SU(3)$ symmetry. From the point of view of the maximally supersymmetric theory, the index counts some 1/16-BPS operators. 
We will take the two chemical potentials for the $SU(3)$ charges to vanish. The superconformal index then takes the form
\begin{equation} \label{Gammaf}\cI={1\over N!} (p,p)^{N-1}(q,q)^{N-1}\int\prod_{i=1}^{N-1} {dz_i\over 2\pi i z_i} \prod_{j\neq k} {\Gamma^3((pq)^{1/3} z_j/z_k ; p ; q)\over \Gamma(z_j/z_k;p;q) }~,\end{equation}
where $(a;b)=\prod_{k=0}^\infty (1-ab^k)$ and 
$\Gamma(z,p,q)=\prod_{k,m=0}^\infty {1-p^{k+1}q^{m+1}/z\over 1-p^kq^mz}$.
The integrals over the $z_i$ variables are over the unit circles. The denominator of the integrand can be interpreted as being due to a vector multiplet (in the $\mathcal{N}=1$ language) while the numerator is due to three chiral multiplets with $r=2/3$ in the adjoint representation.

One should interpret the $z_i$ in~\eqref{Gammaf} in terms of the eigenvalues of the holonomy on $S^1$. Since the holonomy is an $SU(N)$ matrix, the eigenvalues live on a circle. The ratio of $\Gamma$ functions is then some potential for the $SU(N)$ matrix. Since each term in the product depends on some ratio $z_j/z_k$ one can write the matrix model as a double-trace random matrix model. 

To simplify the expressions we will take the gauge group to be $\U(N)$ rather than $\SU(N)$. We rewrite the index $\cI$ as an integral over the $\U(N)$ group manifold. One finds that 
\begin{equation}\label{Mmodel} \cI = \int_{U\in \U(N)} [DU] \exp\left({\sum_{n=1}^\infty  {1\over n} a_n (p,q)\Tr U^n \Tr (U^{-n})}\right)~,\end{equation}
where for the maximally supersymmetric theory,  \be \label{anqp} a_n (p,q) = f(p^n,q^n)\ec \ee and 
$$f(p,q) ={2pq-p-q+3(pq)^{1/3}-3(pq)^{2/3}\over (1-p) (1-q)} \ed$$
The invariant integration measure in terms of the eigenvalues is as usual:
$$[DU]={1\over N!} \prod_{i=1}^N{d\theta_i\over 2\pi}\prod_{1\leq i<j\leq N}|e^{i\theta_i}-e^{i\theta_j}|^2~.$$

Finally, since $\prod_{1\leq i<j\leq N}|e^{i\theta_i}-e^{i\theta_j}|^2 = e^{-\sum_{1\leq i<j\leq N} \sum_k {1\over k}e^{ik\theta_j-ik\theta_i} +c.c.}$ and since we can write $\Tr U^k \Tr (U^{-k}) = \sum_{1\leq i,j\leq N} e^{ik\theta_i-ik\theta_j}=\sum_{1\leq i<j\leq N} e^{ik\theta_i-ik\theta_j}+c.c.+N$, we see that up to an overall factor which is independent of the chemical potentials $p,q$ we can rewrite the superconformal index as  
\begin{equation}\label{apres} \cI = \int \prod_{i=1}^N{d\theta_i\over 2\pi}\exp\left({\sum_{n=1}^\infty{1\over n}\left( a_n(p,q)- 1\right)\Tr U^n \Tr (U^{-n})}\right)~.\end{equation}
Note that $f(p,q)-1={((pq)^{1/3}-1)^{3}\over (1-p) (1-q)}$, i.e. it nicely factorizes.  

A well known observation~\cite{Kinney:2005ej} is that for real $0<p<1,0<q<1$ the scaling of $\log \cI$ is $\cO(1)$ at large $N$. Indeed, we see that since $f(p,q)-1<0$ for all  
such $0<p<1,0<q<1$ the exponential of~\eqref{apres} has all negative coefficients, which means that the most dominant configuration is the one with $\Tr U^n=0$ (for $n<N$), i.e. the eigenvalues are spread out uniformly on the circle and the center symmetry is unbroken, appropriately for the confined phase. In particular for $p=q=e^{-\beta}$ the free energy remains $\cO(1)$ for all positive $\beta$ which signifies large cancellations in~\eqref{OldCIndex} between the bosons and fermions in~\eqref{OldCIndex}. This is a special case of the asymptotic relation~\cite{DiPietro:2014bca} \begin{equation}\label{OldC}\log \cI\sim {1\over \beta}(a-c)~,\end{equation} (where $a,c$ are the usual conformal anomalies) which in the present case leads to a confined phase since $a=c$ in $\mathcal{N}=4$ supersymmetric Yang-Mills theory.

The cancellations between bosons and fermions for $p=q=e^{-\beta}$ prevent a deconfinement transition. These cancellations are puzzling and disappointing since it has been known for a while that 1/16 BPS black holes exist in AdS$_5$ whose macroscopic Bekenstein-Hawking entropy is of order $N^2$ \cite{Gutowski:2004ez,Gutowski:2004yv,Chong:2005da,Kunduri:2006ek}. The lack of a deconfinement transition in the superconformal index with $p=q=e^{-\beta}$ for real beta is not by itself a contradiction, it just signals massive cancellations between the black hole microstates. Some interesting attempts to understand the black hole microstates despite these cancellations have appeared in~\cite{Berkooz:2006wc,Kim:2006he,Grant:2008sk,Chang:2013fba} and see references therein. From the conformal bootstrap perspective, the extremal black hole microstates are some large charge operators. Unlike more familiar large charge limits of the conformal bootstrap~\cite{Hellerman:2015nra,Alvarez-Gaume:2016vff,Monin:2016jmo,Hellerman:2017sur,Jafferis:2017zna,Hellerman:2018xpi,Beccaria:2018xxl,Grassi:2019txd,Sharon:2020mjs,Gaume:2020bmp} where the ground state has $\cO(1)$ entropy, the existence of supersymmetric extremal black holes allows for a large ground state entropy, which is exactly what the superconformal index counts (with various phases). It would be nice to understand if there is a useful explicit effective theory for these states.

More recently, starting with the work of~\cite{CaboBizet:2018ehj,Choi:2018hmj,Choi:2018vbz}, it was understood how to sidestep these enormous cancellations. This gives us a glimpse into the enumeration of black hole microstates. Our approach in this paper is very much based on the random matrix model formalism~\eqref{apres}. A different, very interesting, formalism based on the Bethe equations was developed and used in~\cite{Closset:2017zgf,Benini:2018mlo,Benini:2018ywd}  (see also~\cite{Hosseini:2016cyf,Hong:2018viz,Lezcano:2019pae,Lanir:2019abx}).  We will not have much new to say about this other formalism though we will make some comparisons in section 4.
Some additional references on the subject are~\cite{Honda:2019cio,ArabiArdehali:2019tdm,Cabo-Bizet:2019osg,Amariti:2019mgp,Cabo-Bizet:2019eaf, ArabiArdehali:2019orz,Cabo-Bizet:2020nkr, Murthy:2020rbd,Agarwal:2020zwm,Benini:2020gjh,GonzalezLezcano:2020yeb}. A nice review of the state of the art of black hole microstate counting can be found in~\cite{Zaffaroni:2019dhb}.

Let us describe the idea of how to avoid the cancellations in the superconformal index in general and then specialize to $\mathcal{N}=4$ theory. One considers the index~\eqref{Index} starting from real $p,q$ and then one takes $\tilde p=p e^{2\pi i}$ while $q$ is fixed. In the convention that $p=e^{-\beta+i\sigma_1}$, $q=e^{-\beta+i\sigma_2}$ (where $|p|=|q|$ is assumed), this is the same as shifting $\sigma_1$ by $2\pi$ which corresponds to a particular fibration of the $S^1$ over $S^3$. This transformation multiplies the superconformal index by the phase $e^{-{1\over 3}2\pi i(\Delta+j_1)-2\pi i j_2}$. Using the BPS condition $\Delta=2j_1-{3\over 2}r$ we find that the phase becomes $e^{-2\pi i(j_1+j_2)+\pi i r}$. For fermions $j_1+j_2$ is half integral while for bosons it is integral. Therefore the index turns into 
\begin{equation}\label{IndexN}\mathcal{I} = \Tr_{\mathcal{H}_{\BPS}} e^{\pi i r} \tilde p^{{1\over 3} (\Delta+j_1)+j_2}q^{{1\over 3} (\Delta+j_1)-j_2}~. \end{equation}
The sign $(-1)^F$ is thus replaced by $e^{\pi i r}$ which does not lead to such strong cancellations between BPS states.\footnote{The replacement corresponds to a choice of spin structure on the supersymmetric $S^3\times S^1$ background. In three dimensions this was analyzed in \cite{Closset:2018ghr}.} We can now set $\tilde p=q=e^{-\tilde \beta}$ and take the $\tilde\beta\to 0$ limit which leads to~\cite{Kim:2019yrz}
\begin{equation}\label{NewC}\log \cI \sim {i\over \tilde \beta^2}(3c-2a)~.\end{equation}
An interesting point is that even for theories with $a\neq c$, the growth of states in~\eqref{NewC} is faster than in~\eqref{OldC} since the density of states in~\eqref{OldC} looks like that of a two-dimensional theory while in~\eqref{NewC} it looks like that of a three-dimensional theory.

There is however a very important subtlety here. In~\eqref{NewC} there is an $i$ in the numerator. Therefore, if we strictly set  $\tilde p=q=e^{-\tilde \beta}$ with real $\tilde\beta$ and take the limit $\tilde\beta\to 0$ we would not encounter an exponential growth of states but rather rapid phase oscillations. We should therefore approach the point $\tilde p=q=1$ more carefully, with some imaginary part for $\tilde\beta$. We can only trust the estimate~\eqref{NewC} if the real part of $\log \cI$ is positive. Locally, this means that there is a deconfinement behavior only if one approaches  $\tilde\beta=0$ from one quadrant in the half-plane $\Re \tilde\beta>0$. 

It is straightforward to switch to the $\tilde p,q$ variables in the matrix model description~\eqref{apres}. One finds \begin{equation}\label{apresN} \cI = \int \prod_{i=1}^N{d\theta_i\over 2\pi}\exp\left({\sum_{n=1}^\infty{1\over n}\left(a_n(e^{-2\pi i }\tilde p,q)-1\right)\Tr U^n \Tr (U^{-n})}\right)~.\end{equation}
with 
$$ a_n (e^{-2\pi i}\tilde p, q) = 1- {(1-e^{-2\pi i n/3}(\tilde pq)^{n/3})^{3}\over (1-\tilde p^n) (1-q^n)}~.$$
If we take $\tilde p, q$ to be both real and smaller than 1, then $Re\ a_n  < 1$ for all $n$ and hence the confined phase with homogeneously spread eigenvalues would dominate. No deconfinement transition occurs. As we take $\tilde p, q\to 1$ however, some $ (a_n -1)$ become purely imaginary (essentially all the $a_n$ for $n$ not divisible by 3). This leads to rapid phase oscillations. But it also means that if we now approach $\tilde p, q= 1$ from a slightly different directions some $\Re\ a_n$ can become greater than 1 and a deconfined phase may appear. This is in agreement with~\eqref{NewC}.

Note that in $\mathcal{N}=4$ the following identity holds for all local operators $e^{3\pi i r}=(-1)^F$. (This is a spin-charge relation.) This means that $e^{\pi i r}=e^{-2\pi i r} (-1)^F$ and hence, for $\mathcal{N}=4$ SYM, we can equally well take $p=q$ and rotate both $p,q$ by $e^{-2\pi i}$. We would like to emphasize that this is equivalent to the index~\eqref{IndexN} only in the particular case of  $\mathcal{N}=4$ SYM. Actually for $\mathcal{N}=4$ SYM the following parameterization is very useful: we can take $p=q=ye^{i\psi}$ and study the phases of the theory as a function of $(y,\psi)$. The ``old'' Cardy limit~\eqref{OldC} corresponds to $(y,\psi)=(1,0)$ (at this point and its neighborhood there is confinement, of course) while the ``new'' Cardy limit~\eqref{NewC} corresponds to $(y,\psi)=(1,2\pi)$. Whether or not there is deconfinement depends on how this point is approached. If one keeps strictly $\psi=2\pi$ then there are wild oscillations. Later we will see that this is the edge of the deconfinement region. More generally one would like to study the phases of the theory as a function of $(y,\psi)$. From the definition~\eqref{Index} and the BPS condition $\Delta=2j_1-{3\over 2}r$ we see that the superconformal index is $6\pi$ periodic in $\psi$. 
We will find that in the fundamental domain where $\psi\in(-6\pi,0]$, there is a deconfinement region for $\psi\in (-4\pi,-3\pi)\cup (-3\pi,-2\pi)$.
The two intervals where deconfinement occurs in the fundamental domain are not really different since the superconformal index is the same up to complex conjugation for these two intervals. At the end-points of the intervals $(-4\pi,-3\pi)\cup (-3\pi,-2\pi)$ for no $y<1$ deconfinement occurs. In the bulk of the intervals, deconfinement occurs for finite $y<1$.

Matrix models of the double trace type as~\eqref{apres} have appeared in various applications, especially in connection with partition functions of weakly coupled Yang-Mills theories or as exact results for supersymmetric theories. One common thread in the literature is that the 
deconfinement transition, i.e.  the transition from eigenvalues being spread on the unit circle to eigenvalues which only have a support on a subset of the unit circle, is related to some of the $\Re \ a_n-1$ changing sign. One would think that at the point that $${\rm sgn}(\Re  a_n-1)$$ changes from being positive to negative, one should discard the confined saddle point with $\Tr U=\Tr U^2=\dots=0$ and switch to a deconfined saddle point where the eigenvalue distribution is not invariant under translations along the circle. Indeed, this seems appealing since for $\Re  a_n-1<0$ (for some $n$) the equally spread eigenvalues distribution develops tachyons. 

This logic is in general incorrect. The point is that for complex $a_n$ there could be large cancellations in such a putative deconfined phase and in fact the transition can be delayed. Moreover, unlike in the case of real $  a_n$, the transition is first order and not second order (more precisely, for real $a_n$, it is weakly first order but there is no phase co-existence and hence it looks like a 2nd order transition \cite{Sundborg:1999ue,Aharony:2003sx,Liu:2004vy}). These facts will be crucial for matching the matrix model predictions with the gravitational story. We will demonstrate this phenomenon of a delayed deconfinement transition with a simple ordinary integral in section 2 and then with a toy matrix model in section 3.

 Indeed, in $\mathcal{N}=4$ SYM, one finds that the condition  $\Re (a_n)-1=0$ disagrees with the results from 1/16 BPS black holes in AdS$_5$. This led to some speculations about the existence of new gravitational saddle points \cite{Choi:2018vbz}. 
In fact, $\Re \ a_n=1$ is the natural condition to impose for transitions that are 2nd order or higher. But we will see that the phase transitions are generically strongly 1st order and they are generally delayed, namely, the confined phase may persist even when $ \Re \ a_n>1$. As we said, this is possible when the coefficients in front of $\Tr U^n \Tr (U^{-n})$ are complex due to massive cancellations in the deconfined phase. Our picture fits much better the expectations from gravity. Indeed, the Hawking-Page transition is first order rather than second order and we will see that we are able to reproduce quite precisely the Hawking-Page transition ``temperature'' from the superconformal index!
Therefore no new unfamiliar gravitational saddle points seem to be required to explain the discrepancy between the Hawking-Page transition temperatures computed in field theory and in gravity. We do not solve the matrix model~\eqref{apres} exactly, of course. Rather, we devise a technique to understand this phenomenon of delayed deconfinement through an exact solution of the model with $a_1$ (and also $a_2$ in section 4) and then performing an expansion in the other coefficients. This brings us remarkably close to the gravitational prediction, as can be seen from Figure \ref{curvecomparison1}, comparing the black dashed and red lines.

The techniques we use can be applied in many recently studied problems of microstate counting and phase diagrams of large $N$ theories, including in non-supersymmetric theories such as in the recent work of~\cite{ChermanUnsal:2019hagedorn}, where the study of the phases of gauge theories away from real temperature was undertaken (building upon the previous literature with real $a_n$, see especially \cite{Sundborg:1999ue,Aharony:2003sx,Aharony:2005bq}). Here we will limit ourselves to elucidating the situation in $\mathcal{N}=4$ theory only. Another problem we leave to the future is the question of partially deconfined phases.

This paper is organized as follows. In section 2 we present  a pedagogical example in the realm of ordinary complex integrals showing that modes with negative mass squared do not necessarily ``condense.'' In section 3 we consider a truncated version of the matrix model~\eqref{apres} with only the coefficient $a_1$ present. This truncated model exhibits a lot of the general features, including delayed deconfinement and the edges of the deconfinement curve. This truncated model is simply the Legendre transform of the Gross-Witten-Wadia model~\cite{Gross:1980he,Wadia:1980GWW}. In section 4 we include the effects of the higher $a_n$'s and show that they are numerically small. We compute the corrections to the deconfinement curve and show a remarkable agreement with the Hawking-Page line in AdS$_5$. Two appendices contain technical details. 

\section{Warm Up: An Ordinary Integral}\label{toy}

We will consider in this section the pedagogical example of the ordinary integral 
$$ \int_{-\infty}^\infty dx\, \exp\left(-N\left({1 \over 2}g x^2+{1\over 4}x^4\right)\right)~.$$
The integral converges for all complex $g$. The integral can be done analytically in terms of Bessel functions. It is not essential to keep $N$ as a parameter, but we think about it as a large positive number to make the presentation clearer.

Let us consider some complex $g=|g|e^{i\theta}$. For $\Re g<0$ it seems favorable for  $x^2$ to be away from the origin, such as to increase the magnitude of the term $e^{-{\half}N g x^2}$. This growth should be only terminated by the term ${N\over 4}x^4$, i.e. at $x^2\sim -\Re g>0$ and hence the integral should be naively exponentially large $\sim e^{{1\over 4}(\Re g)^2N}$.
This analysis is perfectly correct for real negative $g$. But in the general case of complex $g$ it is incorrect due to large cancellations. We will see that the integral is not exponentially large for $-3\pi/4\leq \theta\leq 3\pi /4$. In particular, the integral does not become exponentially large when $\Re g =0$ is satisfied. The integral only becomes exponentially large at $\theta=\pm 3\pi /4$ due to an anti-Stokes line, as we will explain below. This sort of delayed exponential growth due to large cancellations is also important in our matrix model, which is why it is useful to go over this pedagogical example first. 

For any nonzero $g$ there are three non-degenerate saddle points: $$0,\pm \sqrt{-g}~.$$ Since the saddle points $\pm \sqrt{-g}$ are generically away from the contour of integration one has to first understand when they should be taken into account. This amounts to finding the Lefschetz thimbles in the complexified $x$-plane. See \cite{Witten:2010cx} for additional details on the subject. There are four degenerate cases: $\theta=0,\pm i,\pi$. As long as we stay away from these four cases, the  Lefschetz thimbles (and their dual upward paths) are non-degenerate and one has a well defined decomposition of our original contour into  Lefschetz thimbles. Let us denote the three thimbles ${\cal J}_0,{\cal J}_{\pm}$ and their dual upward paths $\cK_0, \cK_{\pm}$. We denote our original integration contour by ${\cal C}_{\mathbb R}$. We will restrict to $\Im g\geq 0 $ without loss of generality.

\begin{figure}[t]
	\centering
	\includegraphics[width=1\linewidth]{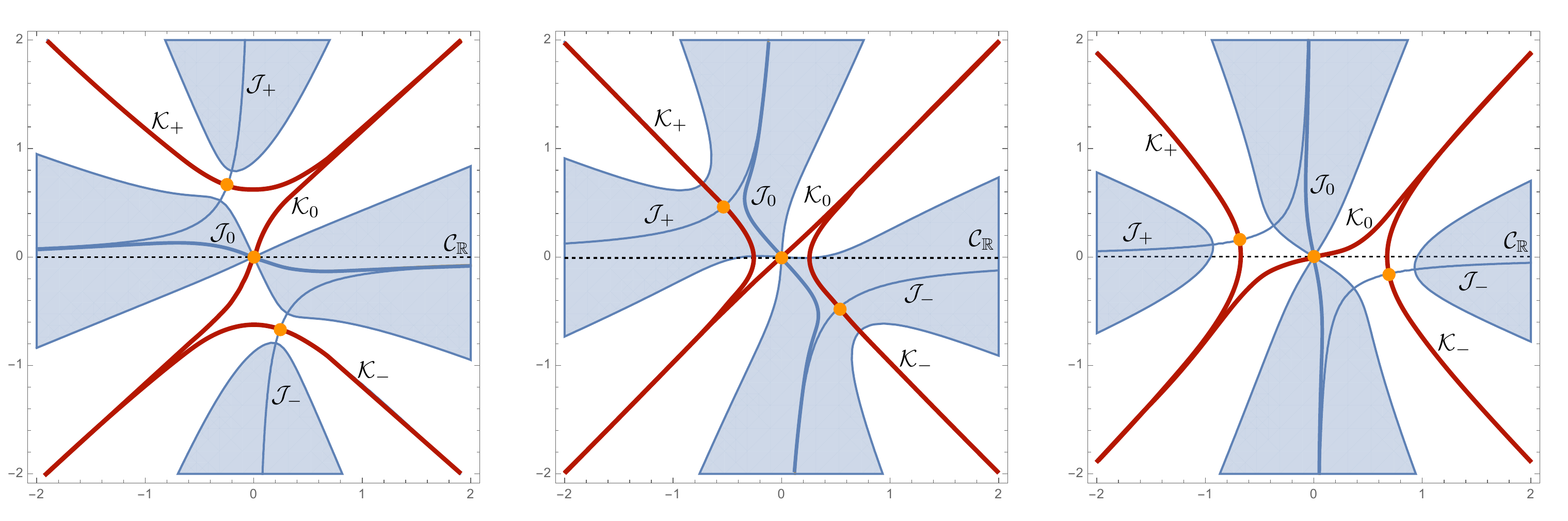}
	\caption{Structure of Lefschetz thimbles (blue color) $\cJ_{0,\pm}$ and upward flow lines (red color) $\cK_{0,\pm}$ respectively for $\theta = 0.7$ (left frame), $\theta = 1.7$ (center frame) and $\theta = 2.7$ (right frame).}
	\label{Thimbles}
\end{figure}

For $0<\theta<\pi/2$ one finds that 
\be\label{firstq}{\cal C}_{\mathbb R} \sim {\cal J}_0~,\ee
where the $\sim$ sign means ${\cal C}_{\mathbb R}$ can be deformed to the thimble on the right hand side. See the left frame of figure \ref{Thimbles}.
Therefore, for $0<\theta<\pi/2$ the saddle points at $\pm \sqrt{-g}$ do not contribute at all. The integral is $\cO(1)$. At $\theta=0$ the upward flow path $\cK_0$ intersects the saddle point at $x= \pm \sqrt{-g}$ and so the situation is somewhat degenerate. But deforming away from this point clearly shows that the ${\cal J}_{\pm}$ do not contribute and the conclusion~\eqref{firstq} therefore also stands for $\theta=0$. 

For $\pi/2<\theta <\pi$ the situation is quite more interesting.
One finds 
\be\label{secondq}{\cal C}_{\mathbb R} \sim {\cal J}_+-{\cal J}_0+{\cal J}_- \ed\ee
See the center frame of figure \ref{Thimbles}. The decomposition has therefore jumped discontinuously at $\theta=\pi/2$, which means that purely imaginary $g$ corresponds to a Stokes line (where the upward flows from the saddle points $\pm \sqrt{-g}$  cross the origin). There is now a contribution from the complex saddle points at $\pm \sqrt{-g}$ which is of the order $\sim e^{{1\over 4} g^2N }$. This is exponentially larger than the $\cO(1)$ contribution from the saddle point at the origin when $\Re g^2 >0$, which corresponds to $3\pi/4<\theta<\pi$. See the right frame of figure \ref{Thimbles}. We therefore identify $\theta=3\pi/4$  as an anti-Stokes line since this complexified saddle point begins to dominate the integral.

Therefore, for $\pi/2<\theta<3\pi/4$ we have $\Re g<0$ and $x$ appears to be tachyonic, but the integral is still $\cO(1)$ due to large cancellations. 
The situation of $\theta=\pi$ is again somewhat degenerate, but the integral is of course exponentially large for $\theta=\pi$  and behaves like $\sim e^{{1\over 4} g^2N }$.\footnote{The degeneracy at $\theta=\pi$ is actually quite more interesting than the one at $\theta=0$ because the decomposition jumps discontinuously for $\theta>\pi$. See, for example,~\cite{Serone:2017nmd} for a discussion of the consequences.} For $\theta=\pi$ all the saddle points are on the real axis and there are no cancellations, which is why the naive estimate gives the correct answer. 

The main take-away lessons from this example are twofold:
\begin{itemize}
\item When the quadratic piece in the action is ``tachyonic'' it does not necessarily mean that the tachyon is about to condense. A phase transition can be delayed if the action is complex.
\item Complexified saddle points may or may not contribute to the integral -- this depends on the structure of the thimbles.
\end{itemize}

\section{Phase Transition at Complex Couplings}\label{cpxhp}
The large-$N$ analysis of the full $\cN=4$ superconformal index  \eqref{Mmodel}, \eqref{anqp}  is quite complicated due to the presence of an infinite number of double-trace interactions. It is therefore instructive to start from the toy  model
\begin{equation}\label{trunc}Z[a_1,N] = \int [DU] \exp\left(a_1 \Tr U \Tr U^{-1}\right)\ec\end{equation}
which  is obtained by truncating \eqref{Mmodel} to the first term.  We refer to \eqref{trunc}  as the $a_1$ model. While we refer to it as a ``toy model,'' it is not actually a trivial system to analyze at all and there is much that is not known about it. 

In the case of real fugacities,\footnote{Namely for real $p, q$ on the first sheet in \eqref{anqp}, such that $a_n$ are all real.} it was shown in \cite{Sundborg:1999ue, Aharony:2003sx, AlvarezGaume:2005fv, Liu:2004vy} that such a toy model captures the essential physics of the problem. In particular \eqref{trunc} gives an exact prediction for the critical parameters where a phase transition of the full model occurs~\eqref{Mmodel}.  The reason is  that the phase transition for real couplings belongs to the second order universality class (more precisely: weakly first order, but since there is no phase coexistence for real $a_1$, it behaves as a second order transition for our purposes) and as in Wilson's general paradigm, the terms corresponding to $a_n$ with $n>1$ are essentially irrelevant.
However, when the couplings are complex, this is no longer the case: the analysis of the double-trace model is modified and many new features emerge. Most notably, as expected also from the gravitational counterpart of this problem, the phase transition becomes (strongly) first order.

The goal of this section is to introduce the large-$N$ analysis for \eqref{trunc} when  $a_1$  takes complex values and to study how this modifies the phase transition observed in the real case.
For the moment, we do not attempt at motivating whether or not the truncation \eqref{trunc} is physically justified. We will address this point in Sections \ref{n4sym} and \ref{twoterms}. 
We  consider the complex model  \eqref{trunc} with $a_1 = |a_1|e^{i \phi}$. In addition, we require ${ \Re(a_1) >  0}$ for the integrals below to converge. At this stage, it is very convenient to introduce the so-called Hubbard-Stratonovich (HS) transformation \cite{Klebanov:1994kv, Liu:2004vy, AlvarezGaume:2005fv}, which allows us to recast \eqref{trunc} as

\be\label{HStrans}
Z[a_1, N] = \int_{\gamma(\phi)} dg g \exp\left(-{N^2 g^2\over 4a_1} \right)\int [DU] \exp\left({Ng \over 2}\left(\Tr U + \Tr U^{-1}\right)\right)\ec
\ee
with $g$ being a complex variable and the integration range of $g$ over $\gamma(\phi)$ is from $0$ to $\infty$ on a line in the complex plane emanating from the origin all the way to infinity at angle $\phi$. 
The introduction of a Hubbard-Stratonovich field $g$ allow us to decouple the double-trace interaction.\footnote{To derive this Hubbard-Stratonovich transformation we start from two conjugate variables $(g, \bar{g})$ and then using the freedom in absorbing the phase of $g$ into the matrix $U$, we can convert the integral over $(g, \bar{g})$ to a radial integral, as in~\eqref{HStrans}  }  As a result, the main problem is now divided into two steps. We first analyze the matrix integral
\be\label{GWW}
Z[g, N] = \int [DU] \exp\left({Ng \over 2}(\Tr U + \Tr U^{-1})\right)\ec
\ee
which is a famous single-trace unitary matrix model whose real-$g$ large-$N$ study was pioneered in \cite{Gross:1980he,Wadia:1980GWW}, it is usually called the Gross-Witten-Wadia (GWW) model. Later we will perform a saddle point analysis of the integral over $g$ in \eqref{HStrans}.

\subsection{The Gross--Witten--Wadia Model }
Let us expand the discussion on the matrix integral \eqref{GWW}. We first 
review some well-known fact about {\it real coupling} $g$. 
It is useful to write \eqref{GWW} in term of eigenvalues\footnote{Alternatively, one can use complex eigenvalues $z_i=e^{i \alpha_i}$:
	\be
	Z[g, N] = \frac{1}{N!} \prod_{i=1}^N \int_{S^1} d z_i \Delta^2(z_i-z_j) \exp\left[N \sum_{i=1}^N\left( \frac{g}{2} \left(z_i + z_i^{-1}\right) - \log(z_i)\right)\right] \ed
	\ee
} $\alpha_i$ of $U=\mathrm{diag}(e^{i\alpha_1},..., e^{i\alpha_N})$ as
\be \label{GWWalpha} Z[g, N] = \frac{1}{N!} \prod_{i=1}^N \int_0^{2\pi}\frac{d \alpha_i}{2 \pi} | \Delta^2(e^{i\alpha_i}-e^{i\alpha_j})| \exp\left(g N \sum_{i=1}^N \cos(\alpha_i)\right) \ed  \ee
Let us first consider $0<g\ll1$. The dominant effect comes from the eigenvalue repulsion in the Vandermonde determinant and consequently the eigenvalues tend to spread out on the circle. Indeed, it is possible to show that \cite{Gross:1980he,Wadia:1980GWW} at large $N$ and $0<g<1$, the eigenvalues are distributed along the whole unit circle with the following density
\be \rho(\alpha)= \frac{1}{2\pi} \left( 1 + g \cos(\alpha) \right)\ec \qquad {\alpha \in [0, 2\pi]} \ed\ee
This makes sense for $0<g<1$ since the density is everywhere non-negative. For this reason this is usually called the ``ungapped phase.''
If $g>1$ the above density is no longer positive definite, hence such a distribution does not make sense.

Indeed at $g=1$ the model has a third order phase transition. For $g>1$ the eigenvalue distribution develops a gap on the unit circle.  More precisely  we have
\be \label{gap} \rho(\alpha) =\frac{g}{\pi} \cos(\alpha/2) \sqrt{\frac{1}{g}-\sin^2(\alpha/2)} \ec \qquad {\alpha\in [-\alpha_c, \alpha_c]}\ed\ee
with {$\alpha_c=2\sin^{-1}\left({1\over \sqrt{g}}\right)$.}

Since in both cases the eigenvalue distribution is supported on one single  interval we refer to such solutions as one-cut. 
The free energy\footnote{Defined as {$F_0(g)=\lim_{N\to \infty}N^{-2}\log Z[N,g]$}.} for these two phases has been computed in \cite{Gross:1980he,Wadia:1980GWW} and reads
\be\label{fzero}
F_0(g) =   \begin{cases}
	{g^2\over 4} , & \quad  \textrm{ungapped, one-cut}  \\
	g -{1\over 2}\log g -{3\over 4}\ec &\quad \textrm{gapped, one-cut} \ed
\end{cases}
\ee
We emphasize that, in the large-$N$ solution of the GWW model, there is no competition between the gapped and ungapped phase since, for real $g$, these solutions do not have an overlapping region of validity. The free energy is a continuous and differentiable function of $g$ for real $g$. There is a third-order transition at $g=1$.

When we turn to generic \emph{complex} $g$ the picture is richer (and not fully understood). Indeed the model also admits multi-cut phases where the eigenvalue distribution is supported on disconnected intervals. Some related results have appeared in \cite{Alvarez:2016rmo}.\footnote{We have computed the two-cut free energy in appendix \ref{2cutApp}. However, as we will see below, this phase is not directly relevant for the physical application we are interested in.}
As we will see below, such multi-cut phases are always suppressed for real $g$. A general way to think about the large-$N$ matrix integral with  $s$-cuts is  by introducing the notion of filling fractions $n_i$. These are defined as
\be \ba  &n_i={N_i\over N} ,  \qquad
&\sum_{i=1}^s N_s=N , \ea\ee
where we can think of $N_i$ as the number of eigenvalues accumulating around the $i^{\rm th}$ cut. The full partition function is schematically  obtained by summing over all possible  arrangements of eigenvalues:
\be\label{general2cut}
Z[g, N] \sim \sum_{N_1 +\cdots+ N_s=N} Z[g, N_1, \cdots, N_s] .
\ee
See for instance \cite{Bonnet:2000dz, Eynard:2008yb, Marino:2008vx} for more details. (It is also possible to add theta-angles for multi-cut phases, but this will not be important here.) In the large-$N$ regime different phases of the model are represented by regions in $g$ space in which a given configuration in \eqref{general2cut} dominates. 
For a given $g$, it is natural to regard the sub-dominant saddles  appearing in the sum as \emph{instanton} solutions which, {when available,} are connected to the dominant configuration by tunnelling processes. These contributions can be estimated explicitly by computing the corresponding instanton action $A(g)$
\be\label{instact}
F^{\textrm{1-inst}}(g) \sim - N^2 A(g)\ed
\ee
Let us now discuss how this picture matches nicely with the analysis  of the  real-$g$  phases discussed above. In particular we want to illustrate how we can explain the GWW phase transition by exploiting the instanton actions. 

We first consider the ungapped one-cut phase.
The instanton action $A^{\textrm{w}}(g)$ was computed and discussed in \cite{Goldschmidt:1979hq, Marino:2008ya} and reads
\be \label{Aw}
A^{\textrm{w}}(g) = 2  \cosh^{-1} \left({1\over g}\right)-2 \sqrt{1-g^2} \ed
\ee
One can think of this action as describing the tunneling from the  ungapped one-cut phase toward a multi-cut phase. See for instance \cite{dunne:2015twocut} for a numerical study and \eqref{aws} for a direct connection with the free energy of the two-cut phase. 

From \eqref{instact} we have that as long as $\Re(A^{\textrm{w}}(g)) >0$, instantons are suppressed and there is no need to consider multi-cut solutions. This is clearly the case if $0<g<1$. 

Similarly,  the instanton action for the gapped one-cut  phase reads \cite{Marino:2008ya}
\be \label{Asdef}
A^{\textrm{s}}(g) = 2\sqrt{g(g-1)} - \cosh^{-1}(2g-1)~.
\ee
One can also think of this action as describing the tunneling from the  gapped one-cut phase towards a multi-cut phase, see for instance \cite{dunne:2015twocut,Alvarez:2016rmo,Marino:2008ya}. In appendix \ref{apfree} we show  that \eqref{Asdef} can actually be written by using the free energy in the the two-cut phase \eqref{aws}.

To have a dominant gapped one-cut  phase we therefore require 
\be\label{1cutdominate}
\Re(A^{\textrm{s}}(g))>0\ed
\ee
This is clearly satisfied for real $g$ with $g>1$.  
Hence, for real $g$, we can explain the phase structure of the GWW model by looking at the instanton actions.

\begin{figure}[t]
	\centering
	\includegraphics[width=0.4\linewidth]{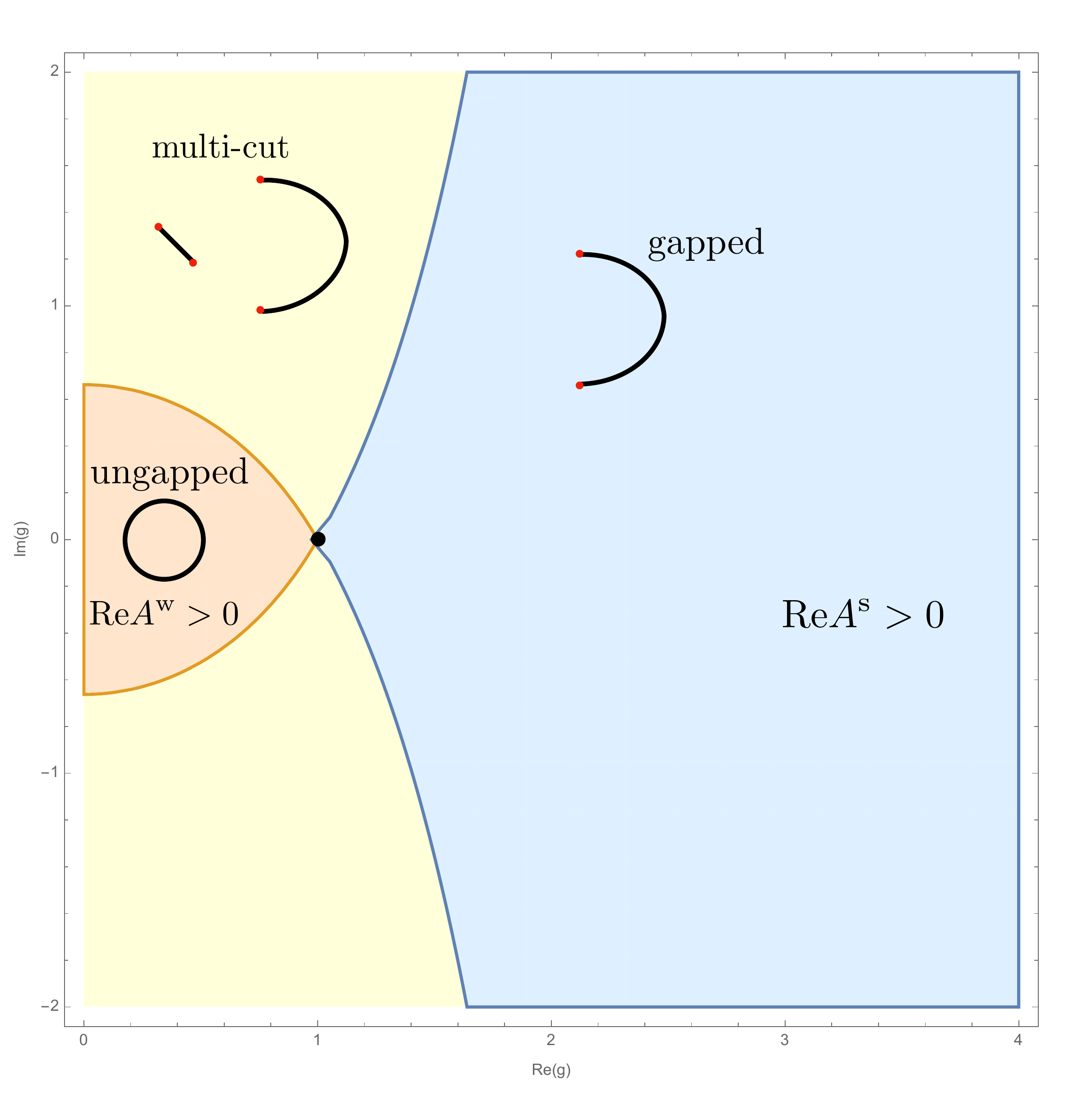}
	\caption{Regions of dominance of various phases in the complex-$g$ plane. The regions are delimited by the anti-Stokes lines $\Re(A^{\ts/\tw}(g))=0$. In the blue region the one-cut gapped phase dominates, in the orange region the ungapped one-cut phase dominates and in the yellow region presumably none of them dominates. While as shown in the appendix one cut phases exist in the yellow ``multi-cut'' region, presumably they do not dominate since the instanton actions for eigenvalues tunneling out of the one cut phases are not suppressed.}
	\label{fig:validity}
\end{figure}

Let us now consider the gapped one-cut phase for complex $g$, without loss of generality we focus on $\Re g > 0$. In particular, we want to know in which region of the $g$-plane the analytic continuation of the gapped solution \eqref{gap} exists and dominates. We expect that this region is determined by the following requirements \cite{David:1990sk}\footnote{See also \cite{Neuberger:1980qh} for an early discussion
and  \cite{Maldacena:2004sn, Marino:2007te,Marino:2008vx, Pasquetti:2009jg,Couso-Santamaria:2015wga} for more recent developments.}
\begin{itemize}
	\item A positive definite analytic continuation of the eigenvalue density \eqref{gap} must exist. 
	\item Instanton tunneling towards others phases should be suppressed. Hence we require: \be \label{a2} \Re (A^s(g))>0. \ee
\end{itemize}
The first requirement is discussed in Appendix \ref{App1} where we argue that 
it is indeed possible to define a positive definite density on the cut for $g \in \C/{[0,1]}$. This follows from a construction of the cut with complexified endpoints $a=1/b=1 - \frac{2}{g} + i \frac{2}{g} \sqrt{g -1}$ that we carry out explicitly. It follows that the only non-trivial\footnote{Notice that the one-cut gapped solution can in principle contribute  also in the region where  $\Re(A^s(g))< 0$. It is possible that somewhere a Stokes line emerges implying that the gapped saddle needs does not contribute. This is exactly analogous to the situation in the toy model of the previous section for real positive $g$, for instance. We have checked that this is indeed the case for the one-instanton solution by imposing $\Im(A^{s/w}(g))=0$. This discussion is not important for our analysis.} constraints is \eqref{a2} which is analyzed in figure \ref{fig:validity}. 

A similar reasoning can be repeated for the weak coupling phase by using $A^\tw(g)$. We thus have analytic control over the two regions in the complex $g$ plane where one-cut solutions exist and dominate shown in Figure \ref{fig:validity}. In the complementary region multi-cut solutions will dominate and in general exhibit a complicated pattern of transitions between them. Luckily, these regions will not be of relevance for our analysis of the confinement/deconfinement transition, as we will conjecture below.

\subsection{Complex Saddles}\label{cpxsaddles}
Our strategy to solve the large-$N$ double trace model \eqref{trunc} consists in performing a saddle point evaluation of the integral \eqref{HStrans} in the variable $g$. 
The same strategy was adopted for example in \cite{Liu:2004vy, AlvarezGaume:2005fv} for real couplings. Nevertheless when $a_1$ is complex, there are some subtle aspects that need to be treated carefully. And as we have mentioned, for complex $a_1$ the transition is first order rather than second order and hence there are also conceptual differences when compared to the case of real $a_1$. 

In order to find the relevant saddle we thus need to consider the following potential
\be\label{Qcomplex}
Q(a_1, g) = -{g^2 \over 4 a_1} + F_0(g)\ec
\ee
where $F_0(g)$ is the genus zero free energy of \eqref{GWW}.
Then, at large-$N$ the integral \eqref{HStrans} is dominated by
\be \label{approx}Z[a_1,N]\approx \sum_{g_*}\exp\left({N^2 Q(a_1, g_*)}\right) \ec \ee
where $Q(a_1, g_*)$ is the potential evaluated over the relevant saddle points $g_*$ of \eqref{HStrans} and will depend on the value of $a_1$. The sum $\sum_{g_*}$ is there to remind us that we have to sum over various saddle points. Since the function $F_{0}(g)$ is naturally defined in a piece-wise manner, we find the saddle point equation in each region and, after having checked that the critical point still lies in said region, we choose the one with the largest $Q(a_1, g_*)$. 

\begin{figure}[t]
\center
	\includegraphics[width=0.45\linewidth]{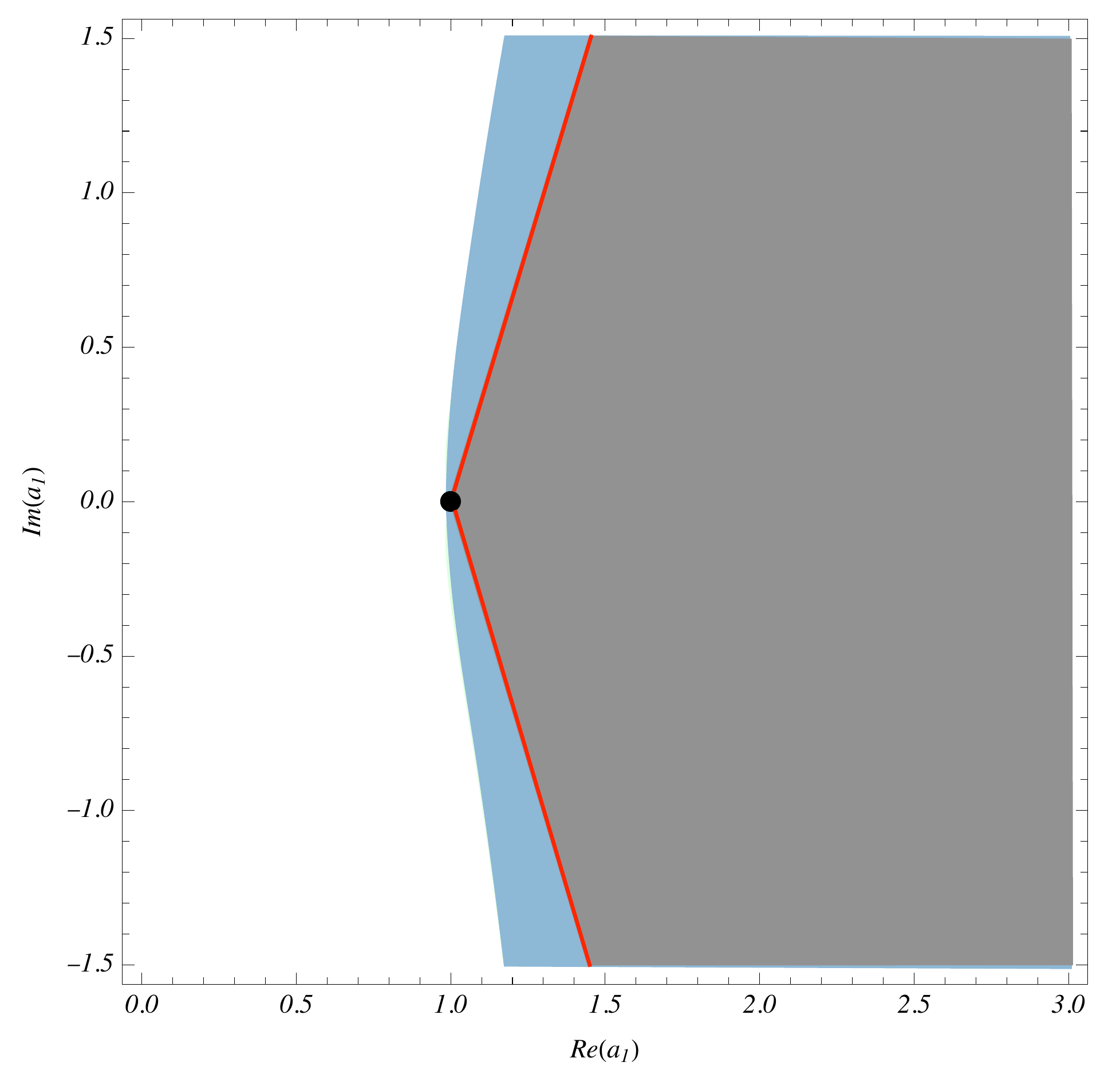}
	\caption{Deconfinement in the complexified model \eqref{trunc}. In both gray and blue areas the gapped one-cut phase contributes as a saddle point. 
	However, only in the grey region we have  deconfinement ($\Re(Q)>0$) . 
	The deconfinement curve $\cC_D$ is drawn in red. The black dot is $a_1=1$.}
	\label{fig:transitiontoy}
\end{figure}

Of course, equation~\eqref{approx} is somewhat loosely written. We need to sum over the saddle points with coefficients depending on the Lefschetz thimbles. In fact, since we do  not have a reliable estimate of  $F_0(g)$ in the whole complex $g$ plane, we are unable to enumerate all the saddle points, leave alone computing the thimbles. Below we will identify two important saddle points which reside in a region in parameter space that we understand in great detail (the blue and orange regions in Figure \ref{fig:transitiontoy}).

Let us first consider the ungapped one-cut phase of \eqref{GWW}. From \eqref{Qcomplex}  and \eqref{fzero} it is clear that $g_*=0$. This corresponds to a saddle point with eigenvalues uniformly spread over the circle. The contribution from this saddle point has no  exponential growth since $\Tr U=0$. 
\be\label{NOgrow}
Z[a_1,N] \sim\mathcal{O}(1)+\sum_{g_*\neq 0}\exp\left({N^2 Q(a_1, g_*)}\right) \ed
\ee

Now let us look for saddle points with $g_*\neq 0$. The free energy of gapped one-cut solutions was given in~\eqref{fzero}. This furnishes a reliable estimate of $F_0(g)$ only for those $g$ that satisfy \eqref{1cutdominate}. In other words, a saddle point from gapped one-cut distributions of eigenvalues would be self-consistent as long as  
\be\label{ags} 
\Re(A^{\textrm{s}}(g_*))>0\ed 
\ee
Using  \eqref{fzero}, we find that the potential \eqref{Qcomplex} has a complex saddle point for \be\label{saddlec}
g_*^{\pm} = a_1 \pm \sqrt{a_1^2-a_1} \ed
\ee
One can verify by explicitly plotting $\Re(A^{\textrm{s}}(g_*^\pm))$ that as long as $\Re(a_1)>1$, the saddle point $g_*^-$ is never in the region where the one-cut phase dominates. This means that only $g_*^+$ potentially falls in the region where we have a self-consistent saddle point with a gapped one-cut distribution of eigenvalues. We hence discard $g_*^-$ and focus our attention on the saddle point $g_*^+$.  We will thus use the notation 
\be \label{g1c}g_*=a_1 + \sqrt{a_1(a_1-1)} \ed \ee
Substituting back we find that $Q(a_1, g_*)$ is given by
\be
\label{qa1}Q(a_1, g_*) = \frac{1}{2} \left(a_1+\sqrt{(a_1-1) a_1}-\log \left(a_1+\sqrt{(a_1-1) a_1}\right)-1\right)\ed
\ee
As a function of complex $a_1$, the potential \eqref{qa1} has a well defined region where 
\be\label{locus}
\Re(Q(a_1, g_*)) > 0 \ed
\ee
Whenever \eqref{locus} is satisfied, the partition function \eqref{HStrans} has exponentially growing behavior.  
On the complex $a_1$-plane, the boundary of \eqref{locus} is given by a complex curve which we denote $\cC_D$ defined by (see Figure \ref{fig:transitiontoy})
\be\label{deccurve}
\cC_D :\, \Re(Q(a_1, g_*)) = 0\ed
\ee
We claim that the curve $\cC_D$ should be used to detect the large-$N$ deconfinement phase transition of \eqref{trunc}.

As we have explained, our computations do not provide a proof that $\cC_D$ is in fact the exact deconfinement curve as not all the saddle points of $Q(a_1, g)$ are presently known. However, $\cC_D$ does agree with the known results for real $a_1$ and it also provides encouraging results in the comparison with gravity, as we will later see.

Let us now make some comments regarding the nature of the phase transition at $\cC_D$ and its order. In the blue region in Figure \ref{fig:transitiontoy} the ungapped one-cut (homogeneously distributed eigenvalues) and the gapped one-cut phase, of the full model, co-exist. While one is in the blue region the ungapped phase dominates and as soon as one enters the grey area the gapped phase begins to dominate. This is clearly a first order transition. For real $a_1$, since the blue region pinches at $a_1=1$, the transition is higher order as the two minima never coexist. Indeed, for real $a_1$, if one approaches from the gapped one-cut phase one has $Z\sim e^{N^2Q(a_1,g_*)}$, with $Q(a_1,g_*)$ given in \eqref{qa1}. For real $a_1>1$, as we approach $a_1=1$, $Q(a_1,g_*)$ goes to zero, and in addition, ${d\over d a_1}Q(a_1,g_*)$ stays finite. In terms of $a_1$ this is a weakly first-order transition. Since for real $a_1$ the two phases never co-exist it is indeed a higher-order transition than for generic complex $a_1$.

\begin{figure}[t]
	\centering
	\includegraphics[width=0.47\linewidth]{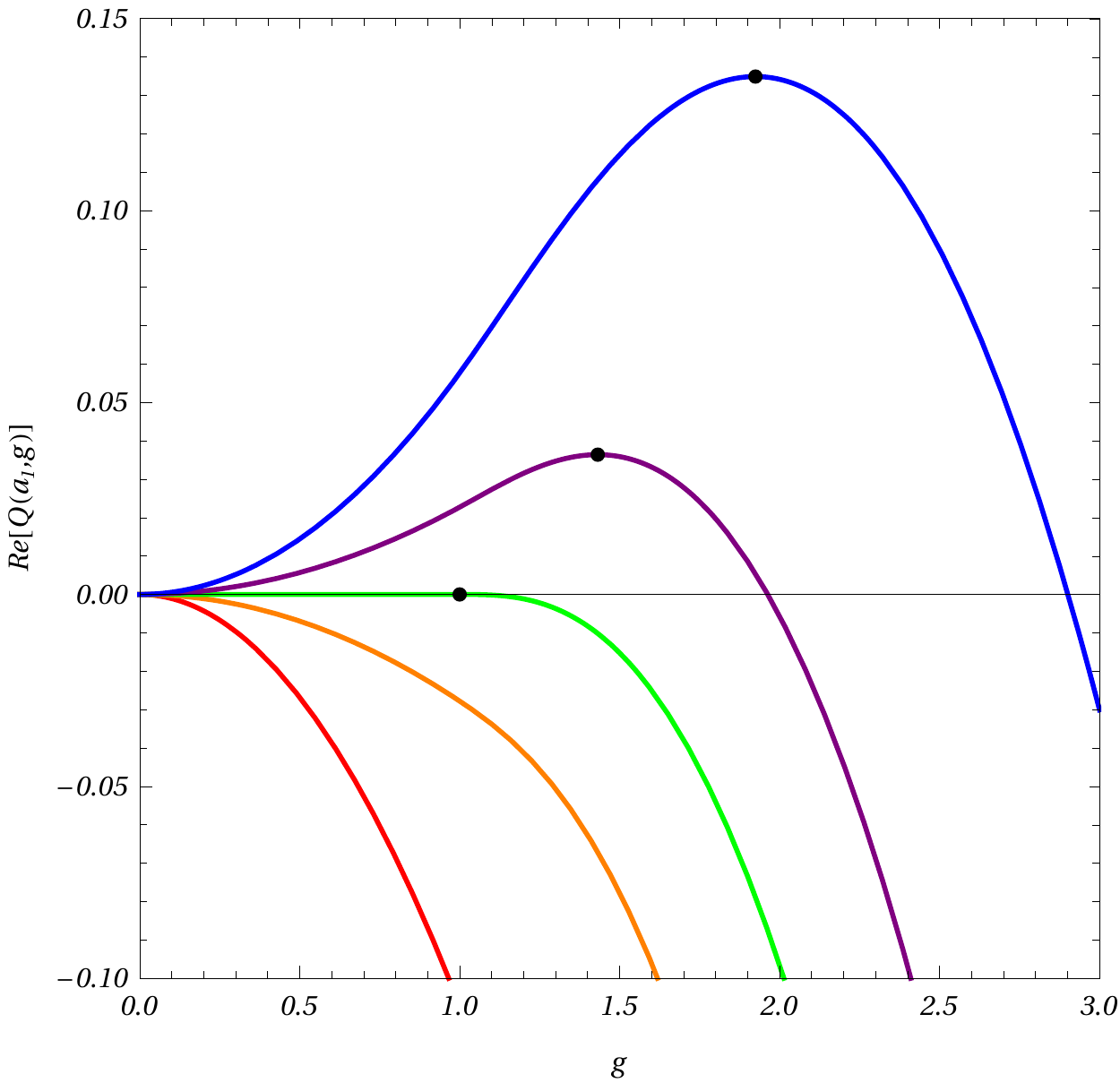} \ \ \includegraphics[width=0.47\linewidth]{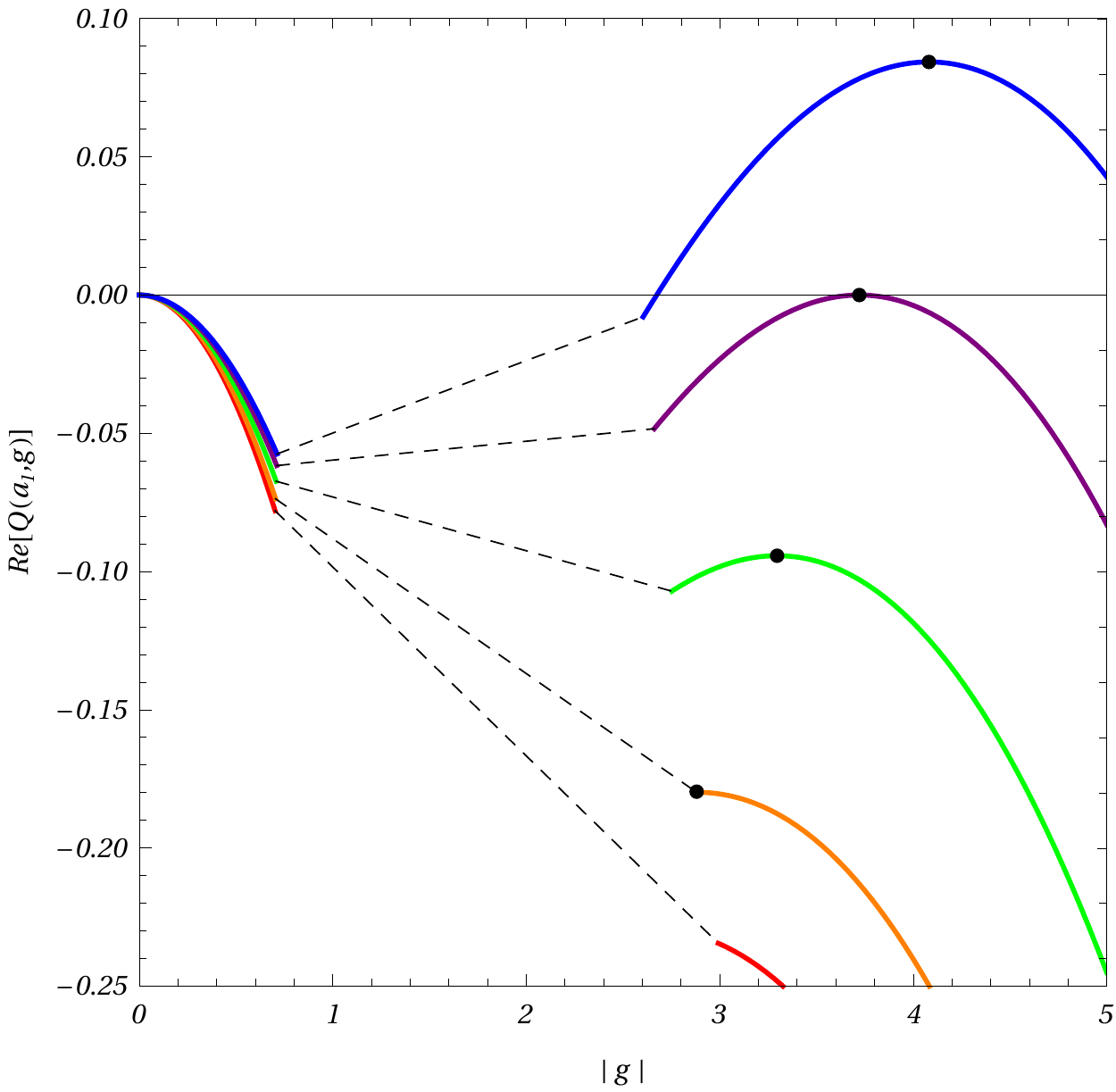}
	\caption{ A plot of \eqref{Qcomplex} in the real (left) and complex case (right, we take $\Arg(a_1)=\pi/4$ for the sake of illustration). Different colors correspond to different values of $|a_1|$. In the second plot, the phase of $g$ is always chosen to be that of $g_*(a_1)$ in \eqref{g1c}.
  Notice that in the complex case the appearance of a second maximum does not coincide with deconfinement. The latter happens only when the second maximum becomes positive. The intermediate dashed region is presumably dominated by multi-cut configurations.}
	\label{fig:realtransition}
\end{figure}

Two comments summarizing our claims:
\begin{itemize} 
	\item For real $a_1$  the integral in \eqref{HStrans} runs over the real line and it is easy to check that \eqref{deccurve} is equivalent to the well known condition 
	\be \label{real} 
	a_1= 1 \ec
	\ee 
	that has been used extensively to detect large-$N$ phase transitions in unitary matrix models with real couplings. This is clearly visible in Figure \ref{fig:transitiontoy}. However, for generic $a_1$ the transition happens beyond the point $\Re(a_1) \geq 1$, i.e. inside the region bounded by $\cC_D$. This means that, even if there are exponentially growing configurations in the original matrix integral ($\Re(a_1)>1$), they will not lead to deconfinement due to destructive interference, as in the toy model in the previous section. 
		\item
	As we discussed around equation \eqref{NOgrow}, the ungapped one-cut phase has a dominating saddle for $g_* =0$. Hence a phase transition can only occur whenever a second minimum of $Q(a_1, g)$ descends below zero (see Figure \ref{fig:realtransition}). Thus, the unitary matrix model with complex coupling  \eqref{Mmodel} has a strongly first order phase transition. This is very different from the physics of unitary matrix models with real couplings which have been used in the past to analyze higher order phase transition. It will have crucial consequences for the validity of the toy model~\eqref{trunc} in the context of the full superconformal index. \end{itemize}

\subsection{The \texorpdfstring{$\cN=4$}{N=4} Supersymmetric Yang--Mills Theory Matrix Model}\label{n4sym}
Let us now connect our results to the analysis of the superconformal index of $\cN=4$ Yang-Mills theory. Following the introduction, we consider the double trace matrix model \eqref{apres}, further specializing the computation to the case of equal fugacities $p=q$. It is also useful to introduce the polar decomposition $p = y e^{i \psi}$. The couplings $a_n$ appearing in \eqref{Mmodel} can be expressed as
\be
a_n(p) =  1 - {(1-p^{2n/3})^3\over (1-p^{n})^2}\ed
\ee
In particular, for $a_1$ we have
\be\label{aone}
a_1(p) = 1 - {(1-p^{2/3})^3\over (1-p)^2}\ed
\ee
Using the explicit form of $a_1$ we can now apply the results from section \ref{cpxsaddles}. As before, we ignore for the moment the contributions of the higher $a_n$. We will justify this in the next section.

With the previous discussion in mind, we define a function
\be\label{ourf}
f(p) \equiv \Re(Q(a_1(p),g_*))\ec
\ee
given by the potential \eqref{Qcomplex} evaluated in the gapped one-cut phase and on the complex saddle \eqref{saddlec} with $a_1$ expressed as in \eqref{aone}. We expect this to give a good first approximation for the phases of the superconformal index.

\begin{figure}
	\centering
	\includegraphics[width=0.8\linewidth]{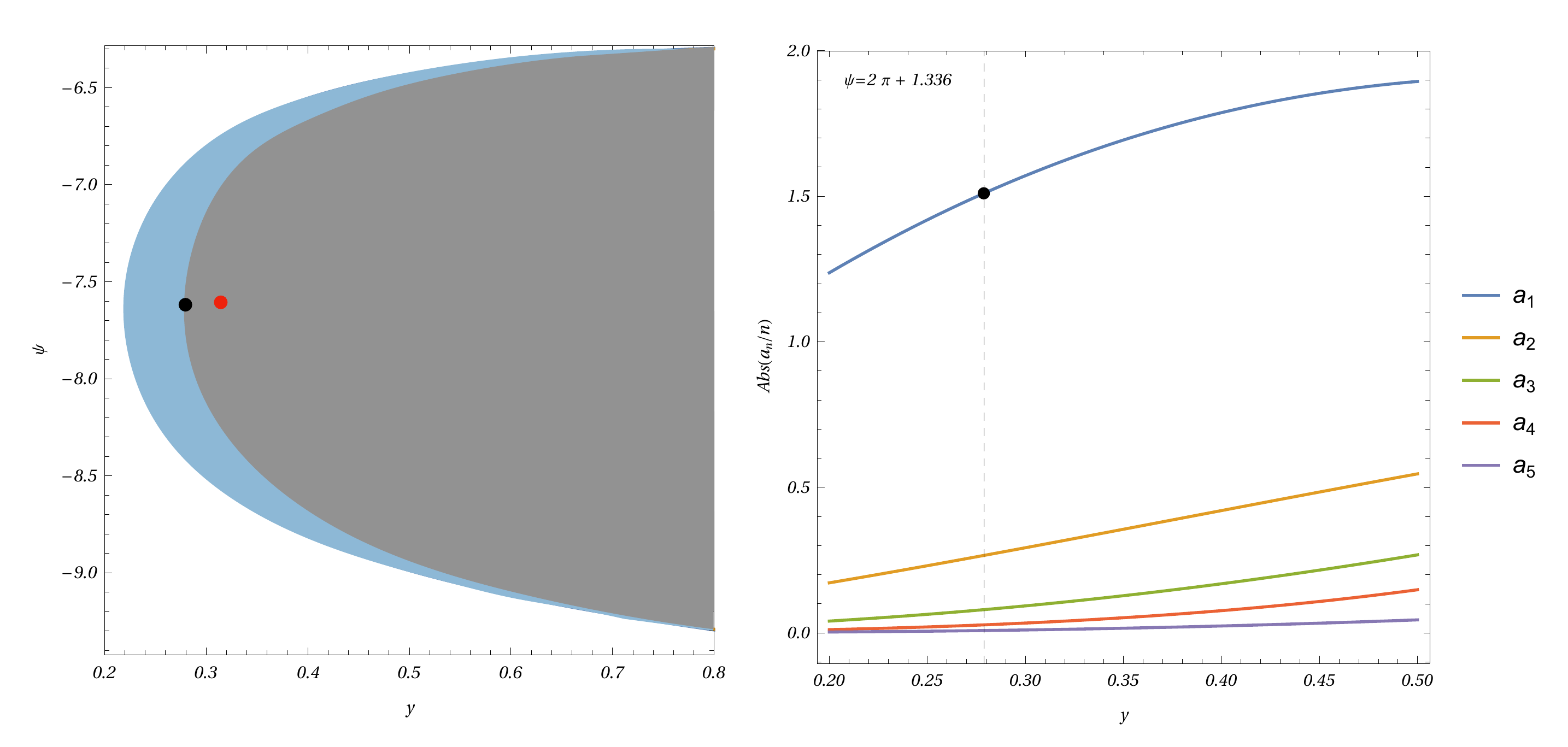}
	\caption{Left: Deconfinement for the truncated model in the $\mathcal{N}=4$ case. The grey area is the prediction coming from our analysis, while the blue area is the region where tachyonic modes exist ($\Re a_1 >1$) but the index is not yet deconfined. The black dot is the point $p_\star$ while the red point is the gravity prediction $p_{\text{HP}}$. Right: The numerical values of higher $a_n$ as a function of $y$.}
	\label{fig:transition}
\end{figure}

It is easy to plot the function $f$ in the $(y,\psi)$-plane, as we did in Figure \ref{fig:transition}. Clearly, the locus where $f$ is zero on the $(y,\psi)$-plane defines the deconfining curve $\cC_D$. This means that there exists a range of $(y,\psi)$ values for which $\Re(Q(a_1, g_*))$ satisfies \eqref{locus} and leads to exponential behavior. In this region, the large-$N$ field theory has a \emph{deconfining} behavior with large-$N$ free energy scaling as $\cO(N^2)$ with a positive coefficient. Note that the point $a_1=1$ where the transition in the original model~\eqref{trunc} is higher order has essentially disappeared in the $(y,\psi)$ variables. This is very encouraging, as we  expect the Hawking-Page transition to be always first order in gravity.

We can consider the point $p_{\star} \in \cC_D$ which maximizes the absolute value of $\omega$:
\be\label{maxus}
p_{\star} \approx 0.279\,\re^{ -(2 \pi + 1.336)\, i}\ed
\ee
The very first deconfining phase transition that can be detected on the $(y,\psi)$-plane will occur exactly at this point. The importance of this special point will be discussed in section \ref{hptrans}.

Direct computation shows that, in the neighborhood of $p_{\star}$, instantons are still suppressed, hence we are well within the regime where our analysis is correct:
\be
\Re(A^{\textrm{s}}(g_*(p_\star))) \approx 0.0074~.
\ee 

In the region around $p_\star$, the higher $a_n$ are somewhat numerically small, which explains why our result for $p_{\star}$ agrees quite well with the gravitational computation to follow.\footnote{If one imposes the more naive criterion $Re(a_1)>1$ one finds the maximizing point \cite{Choi:2018hmj, Choi:2018vbz}\be\label{maxkor}
p_{\star}^{\textrm{CKKN}} \approx 0.226\,\re^{-(2\pi + 1.336) \,i}\ec
\ee
which indeed has smaller real  part than \eqref{maxus} and does not agree with the Hawking-Page transition in the bulk.} 
We will compute the corrections from $a_2$ in the next section and we will see that the agreement becomes even better.  

To conclude this section, let us briefly comment on the new Cardy limit (upper--right edge in Figure \ref{fig:transition}) which we described in the introduction. In the new Cardy limit $p=q=e^{-\omega-2\pi i}$, $|\omega|\ll 1$ the index can be resummed explicitly \cite{Choi:2018hmj}. From \eqref{NewC},\footnote{In the variables $y$, $\psi$ this means $y$ slightly smaller than $1$ and $\psi$ smaller than $-2 \pi$}  deconfined behavior is expected for
\be  \Im\left( \omega\right) > 0, \quad \Re \left( \omega\right) \to 0^+ . \ee
In this limit the $a_n$ coefficients behave as
\begin{equation}
\frac{a_n}{n} = - \frac{6 i}{n^3 \omega^2} \sin\left( \frac{4}{3} n \pi \right) + \cO(\omega^{-1}) \, ,
\end{equation}
so that all of the $a_n$'s are parametrically growing and it would seem reckless to consider the model with only $a_1$. Yet, as we can see in Figure \ref{fig:transition}, our model reproduces a similar looking edge, which locally divides the space at $y=1,\psi=-2\pi $ to two half-planes one of which is in the deconfined phase and the other in the confined phase.

Let us demonstrate this analytically in the model with only $a_1$. We expand around
$$a_1 = - \frac{6 i}{ \omega^2} \sin\left( \frac{4}{3} \pi \right)+\cdots$$
This means that \be g_* = 2 a_1 + \cO(1)\ed \ee
The instanton action \eqref{ags} is dominated by the square root part, so \be  A^\ts(g_*) = 4 a_1+\mathcal{O}(\log(a_1))\ed\ee
This implies that instantons are suppressed precisely in the region of $\Im \omega > 0$ and $\Re \omega$ small and positive. In addition  the potential
\eqref{Qcomplex} now reads
\begin{equation}
Q(a_1,g_*) = a_1 + \cO(\log(a_1)) \, .
\end{equation} 
This means that $\log Z\sim {i\over \omega^2}$ near the Cardy limit, in qualitative agreement from what one expects of the full superconformal index~\eqref{NewC}.
Therefore our toy model correctly captures the deconfinement transition of the full index also at the new Cardy limit. 
 
The above discussion on the Cardy limit can be generalized to a limit where only a subset of the $a_k$ coefficients in the matrix model becomes parametrically large. Such limit, which corresponds to $p \to e^{2\pi i {n \over m}}$ from below with $n,m$ co-prime integers, has been analyzed in \cite{Cabo-Bizet:2019eaf}. Following the same logic we adopted in this section we can study a model \be Z[a_m, N] = \int dg_m \,g_m \exp\left(-{N^2 g_m^2\, m\over 4 a_m } \right)\int [DU] \exp\left({Ng_m \over 2}\left(\Tr U^m + \Tr U^{-m}\right)\right)\ed\ee A preliminary test suggests that the above model is dominated by partially deconfined saddles with $\bZ_m$ symmetry. It would be interesting to use this approach to study the nature of the saddles appearing in \cite{Cabo-Bizet:2019eaf} and their putative gravitational duals.

\subsection{Black Hole Entropy and Hawking-Page Phase Transition}\label{hptrans}
In section \ref{cpxsaddles} we have shown that there exists a special point $p_\star \in \cC_D$ where deconfinement is first supposed to take place. As such, it is quite natural to ask whether $p_\star$ might have a physical significance in the context of the AdS/CFT correspondence, to which we alluded many times above. 

It has been appreciated for a long time that the gravitational dual of large-$N$ deconfinement transition should be thought of as a Hawking-Page transition \cite{Witten:1998zw}. At the semi-classical level the gravitational path integral is dominated by saddle point configurations of two kinds, one of them is thermal (Euclidean) AdS$_5$ while the other is a large AdS$_5$ black hole. The large-$N$ free energy contribution from thermal $\AdS_5$ is expected to be $\cO(1)$. This behavior changes when the system reaches a critical Hawking-Page temperature $T_{\HP}$ where the large $\AdS_5$ black hole saddle dominates and the free energy grows as $\cO(N^2)$.  

Here we are interested in ${1/16}$-BPS black holes  solutions of IIB string theory on $\AdS_5\times S^5$ \cite{Gutowski:2004ez,Chong:2005da,Kunduri:2006ek,Chong:2005hr,Cvetic:1999xp}. 
Upon compactification on $S^5$, these are five dimensional asymptotically $\AdS_5$ black holes  whose near horizon geometry consists of an $\AdS_2$ fibration over $S^3$ \cite{Kunduri:2007qy}.
Such solutions are characterized by three electric charges $Q_{1,2,3}$ and two angular momenta $J_{1, 2}$ whose corresponding fugacities are denoted by $\Delta_{1,2,3}$ and $\omega_{1,2}$. 
The five charges $Q_I$, $J_K$ satisfying some highly non-trivial constraints which, at the level of the fugacities, can be simply expressed as (see also~\cite{Larsen:2019oll}) \footnote{For BPS states which have $E \geq \sum_{i=1}^{3}Q_i + \sum^{2}_{a=1}J_a.$}
\be\label{surface}
\Delta_1 + \Delta_2 + \Delta_3 - \omega_1 -\omega_2 =- 2\pi i\ed
\ee
The entropy for these black holes was computed in \cite{Kim:2006he} and reads
\be  \label{sknown}S= 2\pi \sqrt{Q_1 Q_2+Q_1 Q_3+Q_2 Q_3-{\pi \ell_5^3 \over 4 G_N }(J_1+J_2)},\ee
where $\ell_5$ is the $\AdS_5$ radius and $G_N$ is the the five-dimensional Newton constant. 
It was recently shown \cite{Hosseini:2017mds} that \eqref{sknown} is in fact the result of an extremization procedure.  
More precisely we have
\be \label{sext}
S =  -F_{\textrm{BH}} + \sum^3_{i=1} \Delta_i Q_i + \sum^2_{a=1} \omega_a J_a\ec
\ee
where 
\be \label{free}
-F_{\textrm{BH}} = {N^2 \Delta_1\Delta_2\Delta_3 \over 2 \omega_1 \omega_2}\ec
\ee
can be thought as  the leading terms in the large $N$   {black hole} free energy. Moreover one should evaluate \eqref{sext} for fugacities fulfilling the ``extremization constraint"
\be\label{ext}{\partial S \over \partial \Delta_i}=0= {\partial S \over \partial \omega_k}.  \ee
Let us now consider a simpler situation with equal charges $Q_1=Q_2=Q_3= Q$ and equal angular momenta $J_1=J_2=J$. From \eqref{surface}, this also implies that $3\Delta -2\omega = -2\pi i$. Then we have
\be \label{FBH} F_{\textrm{BH}} = -{4 N^2 ({-\pi i +  \omega} )^3 \over 27 \omega ^2}\ec
\ee
where $\omega = -\log y -\ri (2\pi + \psi) $ in the conventions of section \ref{n4sym}. Therefore the Hawking--Page line  where black holes start to dominate over thermal AdS$_5$ is given by
\be \Re(F_{\textrm{BH}})=0.\ee 
Note however that the extremalization procedure will in general lead to complex charges for generic fugacities because of \eqref{surface}. These will not correspond to physical black hole solutions. Since the charges are equal, the reality condition determines a line in the complex fugacity plane.
The intersection between this line and the Hawking-Page line defines a point in the fugacity plane. At this location we both have a physical black hole solution (with real charges) and we cross the Hawking-Page line. We will call such a point the Hawking--Page transition $\omega_{HP}$. 
Converting to our conventions for the fugacities one finds \cite{Choi:2018hmj}\footnote{From now on we identify $p = e^{-\omega}$, this slightly differs from the $\omega_k$ used above by a factor of $2 \pi$. See for example \cite{Benini:2018ywd} for the explicit dictionary. We hope that this will not generate any confusion.}:
\be\label{whp}
\omega_\HP = \frac{\pi}{16} \sqrt{414 - 66\sqrt{33}}  + \ri \,{\pi\over 16}(1 + \sqrt{33}+ 16)\ed
\ee
Which implies:
\be\label{php}
p_{\HP} \approx 0.314 \,\re^{-(2\pi + 1.324)\ri\,}\ed
\ee
If we compare $p_\HP$ with our $p_\star$ we find 
\be\label{discrepancy}
|p_{\star}| < |p_{\HP}|\ed
\ee
See also Figure \ref{fig:transition}. Let us now discuss some consequences of the above result. First, if our $p_\star$ were the true deconfining critical value for the full superconformal index \eqref{Gammaf}, deconfinement would take place in field theory before a known large AdS black hole saddle dominate the gravitational ensemble. In other words, this would necessarily imply that yet unknown gravitational saddles with have to be found in order to save the AdS/CFT correspondence.

However, since the transition at  $p_\star$  was shown to be first order, it may shift due to $a_{n>1}$ corrections. We will see that these corrections can indeed be taken into account, they are numerically somewhat small, and they remove the discrepancy with AdS/CFT. In other words, there is presently no need for new gravitational saddles.

\section{Corrections to the Deconfinement Curve} \label{twoterms}
In section \ref{cpxhp} we analyzed a simple unitary matrix model \eqref{trunc} which is obtained as a truncation of the $\cN=4$ superconformal index matrix model to its first interacting term with coupling $a_1$. 
Since the truncated model has a first order transition for complex $a_1$, there is no sense in which the higher $a_n$ may be ``irrelevant.'' This is in contrast to the case of real $a_1$, where the transition of the truncated model is a higher order one, and hence some of its predictions are exact. 
Therefore, for complex $a_1$, the deconfinement curve may and will receive corrections from the higher $a_n$'s. At this point it may seem hard to obtain a useful approximation of the deconfinement curve. Accounting for the effects of the $a_n$'s  is indeed difficult. However, we will show here that near the deconfinement transition the effects of the higher $a_n$'s are numerically small and going to second order is already sufficient to obtain a remarkably close deconfinement curve to the one predicted by gravity.  

We emphasize again: in the analysis below, there is no parametric suppression of the remaining $a_n$ corrections. In this sense the situation is more akin to perturbative computations in QED, where the electromagnetic coupling just ``happens'' to be sufficiently small to get accurate results by simple perturbation theory.

\subsection{Two-Terms Model}\label{2terms}

The simplest improvement of \eqref{trunc} is obtained by adding the second smallest coupling near the region of the Hawking--Page transition, namely $a_2$. 
\be \label{original} 
Z[a_1, a_2, N] = \int [DU] \exp \left(a_1 \Tr U \Tr U^{-1}+{a_2\over 2} \Tr U^2 \Tr U^{-2} \right)\ed
\ee
We assume that a small enough $a_2$ only changes the original one-cut saddle in a continuous way so that no further transition happens. In fact, we will end up treating $a_2$ perturbatively so this is guaranteed.
To find the corrections to the deconfinement curve it is again convenient to apply the Hubbard-Stratonovich transformation to get
\be\label{HSgen}
\begin{split}
	&Z[a_1, a_2,N] = \int g\, dg\, z_1 z_2  dz_1 dz_2 \exp \left(-{N^2 g^2\over 4 a_1}+{N^2  (z_1^2+z_2^2) \over  {2 a_2}}\right) \times  \\
	&\int [DU]\, \exp\left( {Ng\over 2}(\Tr U+ \Tr U^{-1})+{\ri N z_1  \over 2}\left( \Tr U^2+ \Tr U^{-2}\right)+{\ri N z_2  \over 2}\left( \Tr U^2- \Tr U^{{-2}}\right)\right)\ed
\end{split}
\ee
Here, as before, we have redefined the phase of $U$ in order to have just a single complex coupling $g$ for $\Tr U$. This however still leaves us with three complex couplings, which make the model much harder to analyze. A simplification occurs by noticing that the original model has a  $\bZ_2$ (charge conjugation) symmetry 
\be  U \leftrightarrow  U^{-1}\ed
\ee
under which $z_2$ is an odd coupling. Therefore, if we expand the free energy around $z_2=0$ it will have only even powers (assuming it is analytic) and hence $z_2=0$ is a saddle point of~\eqref{HSgen}.
Thus, as long as we wish to study the leading saddle in the regime where $a_2$ effects are small, we may safely set $z_2=0$. Therefore we wish to calculate the large-$N$ behavior of 
\be \label{HStrans2}
\begin{split}
	\int g\,dg \, h\,dh &\exp\left(-{N^2 g^2\over 4 a_1} -{ N^2  h^2 \over  { 2 a_2}}\right)\times \\  &\int [DU] \exp\left( {Ng\over 2}(\Tr U+ \Tr U^{-1})+{N h\over 2} \left( \Tr U^2+ \Tr U^{-2} \right)\right)\ec
\end{split}
\ee
where we set $ h \equiv \ri z_1  $. Following the same strategy adopted in section \ref{cpxhp} we will first focus on the integral \be \label{zab}
Z[g,h]=\int [DU] \exp \left( {N g\over 2}(\Tr U+ \Tr U^{-1})+{N h\over 2} \left( \Tr U^2+ \Tr U^{-2} \right) \right)\ed
\ee 
More precisely, up to some unimportant normalization factors we write \eqref{zab} as 
\be\label{cpxmodel}   
Z[g,h]= \int_{\Gamma}\rd z^N \prod_{i<j} \left(z_i-z_j\right)^2\prod_i \exp\left[ N V(z_i)\right]\ec
\ee
where 
\be\label{pot} 
V(z)= {g\over 2} (z+z^{-1})+{h\over 2} (z^2+z^{-2})- \log z \ed
\ee
When $g$ and $h$ are both real, the contour $\Gamma$ is just the unit circle since the integral \eqref{original} is performed over unitary matrices. However we would like to
study \eqref{cpxmodel} for $g$ and $h$ both complex.  $\Gamma$ then becomes a more general path in the complex plane. 
The model \eqref{zab} admits multi-cut phases, see for instance \cite{Jurkiewicz:1982iz}. Thus in order to make contact with the Hawking-Page transition described in section \ref{hptrans} we further need 
to restrict our attention to a region of the complex $(g,h)$-parameters space where the one-cut phase is well defined.  In order to compute the genus zero free energy $F_0(g,h)$ it is useful to consult the classic reference 
\cite{Jurkiewicz:1982iz}, see also \cite{Eynard:2015aea} for a pedagogical review.  {We review some aspects of our main computation in Appendix \ref{apfree}.  We find that
the full genus-zero free energy $F_0(g,h)$ in the one cut phase is
\be\label{f0gb}
F_0(g,h)=\frac{1}{4} \left(2 h^2 \Delta ^4+\Delta ^2 (g-4 h (\Delta -1))^2-4 (\Delta -1) (-3 h \Delta +h+g)+2 \log (\Delta )\right)\ec
\ee
where
\be 
\Delta \equiv \frac{2}{4 h+g+\sqrt{(4 h+g)^2-24 h}}\ed
\ee
Hence, at large-$N$, the integral \eqref{HStrans2} can be further simplified to
\be\label{gbfinal}
Z[a_1,a_2, N] \sim \int  dg\, dh \exp\left[-{N^2 g^2\over 4 a_1}-{N^2  h^2 \over  { 2 a_2}}+N^2F_0(g,h)\right]\ec
\ee
with $F_0(g, h)$ given by \eqref{f0gb}. 

Our final task is to perform the large-$N$ saddle point analysis of \eqref{gbfinal}. However, in contrast with section \ref{cpxsaddles}, such analysis is further complicated by the  form of $F_0(g,h)$ which does not allow us to solve the saddle point equations for $(g_*,h_*)$ analytically. This is not a major obstacle since we can now assume that $a_2$ is small, which in turn implies that $h$ is also small on the relevant saddle. Thus we may obtain analytic formulae by simply considering the perturbative regime.
The large-$N$ free energy $F_0(g,h)$ can be expressed as a series in $h/g$ and the saddle point equations for the couplings can be solved recursively.

By performing a perturbative analysis up  second order in $a_2$ we find that the original model \eqref{gbfinal} can be approximated by
\be\label{exptwo}
\exp\left[ -{N^2 (g^{(2)}_*)^2\over 4 a_1}-{ N^2  (h^{(2)}_*)^2 \over { 2 }a_2}+N^2F_0^{(2)}(g^{(2)}_*,h^{(2)}_*)\right] \ed
\ee
where 
\be F_0^{(2)}(g,h)= \frac{1}{4} \left(\log \left(\frac{1}{g^2}\right)+4g-3\right)+\frac{(1-g)^2 h}{g^2}+\frac{(16 (g/2-1) g+9) h^2}{8 g^4}\ed
\ee
and
\begin{align}
h_*^{(2)} & = { \left(1-\frac{1}{a_1}\right)} a_2 \,+\\\nonumber
&+ \frac{ {(a_1-1)} \left(4 a_1 \left(2 a_1-2 \sqrt{(a_1-1) a_1}-3\right)+8 \sqrt{(a_1-1) a_1}+5\right) { a_2^2}}{ a_1^3} \ec\\
g_*^{(2)} &=\left(a_1+\sqrt{(a_1-1)a_1}-\frac{{ 2} (a_1-1) \left(\sqrt{(a_1-1)a_1}-a_1\right)a_2}{a_1^2}\right)\ed
\end{align}
It is easy to iterate this procedure and repeat the analysis to a higher order.  

Note that thus far we have only considered the effects of $a_2$. We are not presenting the results of including the effects of $a_{n>2}$ since these corrections turn out to be much smaller numerically than the $a_2$ corrections. 

\subsection{Comparing with the  Hawking--Page Phase Transition}\label{morehp}
We can now use our analysis of section \ref{2terms} to compare with the gravity prediction described in section \ref{hptrans}. 
Using the explicit expressions for $a_1$ and $a_2$ in \eqref{exptwo}
\be \ba a_1=&1 - {(1-p^{2/3})^3\over (1-p)^2}\ec\\
a_2=& 1 - {(1-p^{4/3})^3\over (1-p^{2})^2}\ec&
\ea\ee
we now study the corresponding deconfining curve as a function of $p=y \re^{\ri \psi} \in \IC$. This gives a curve in the
 $(y,\psi)$ plane which is shown in Figure \ref{curvecomparison1}.
 
 \begin{figure}[t]
	\centering
	\includegraphics[width=0.5\linewidth]{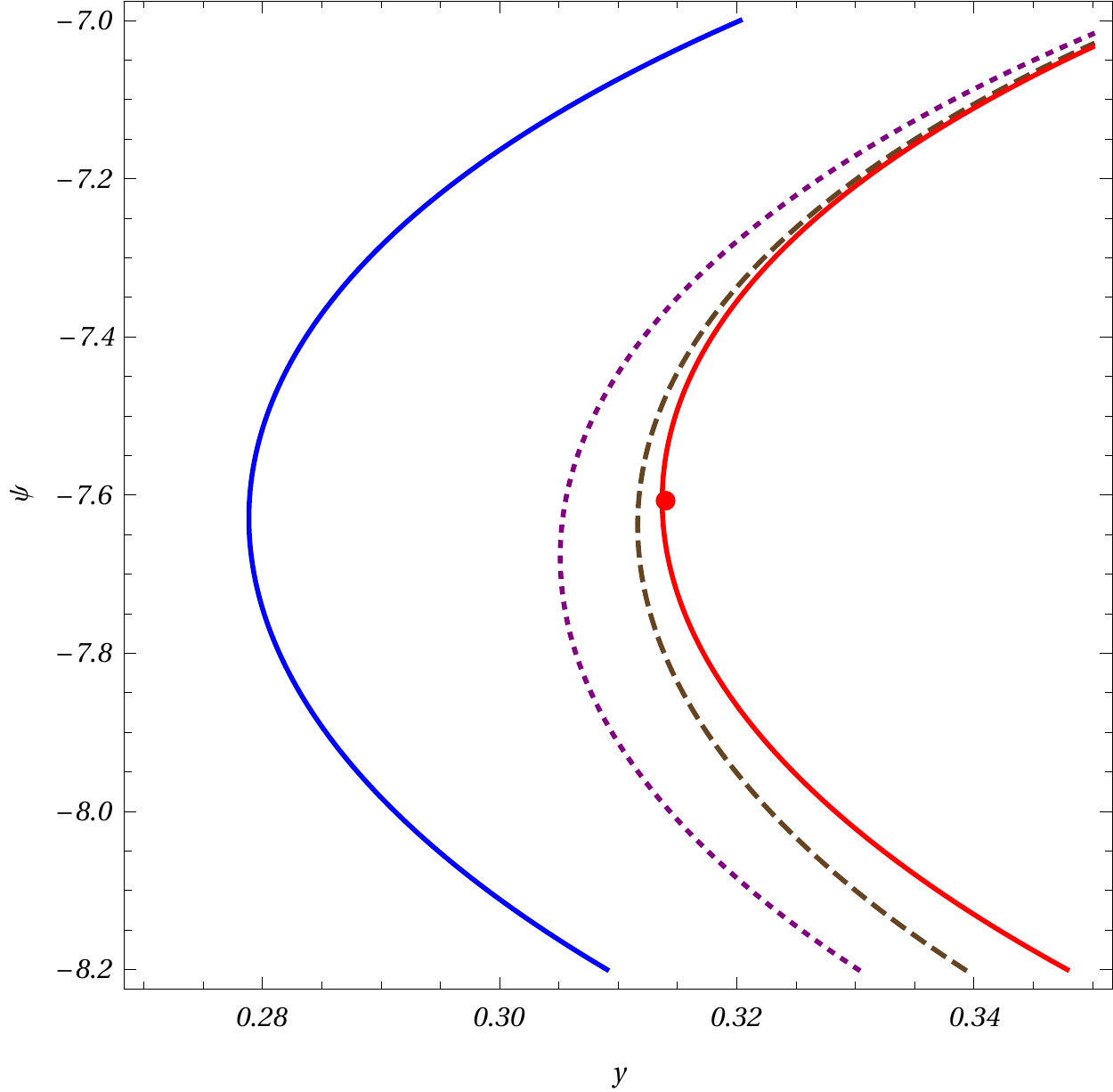}
	\caption{The blue, purple dashed and black dashed curves are respectively deconfinement curves on the $(y, \psi)$-plane for the $a_1$-model and the $\cO(a_2)$, $\cO(a_2^2)$ corrections. The red line is the gravitational curve $\Re F_{\textrm{BH}}=0$.}
	\label{curvecomparison1}
\end{figure}

 \begin{figure}[h]
	\centering
	\includegraphics[width=0.5\linewidth]{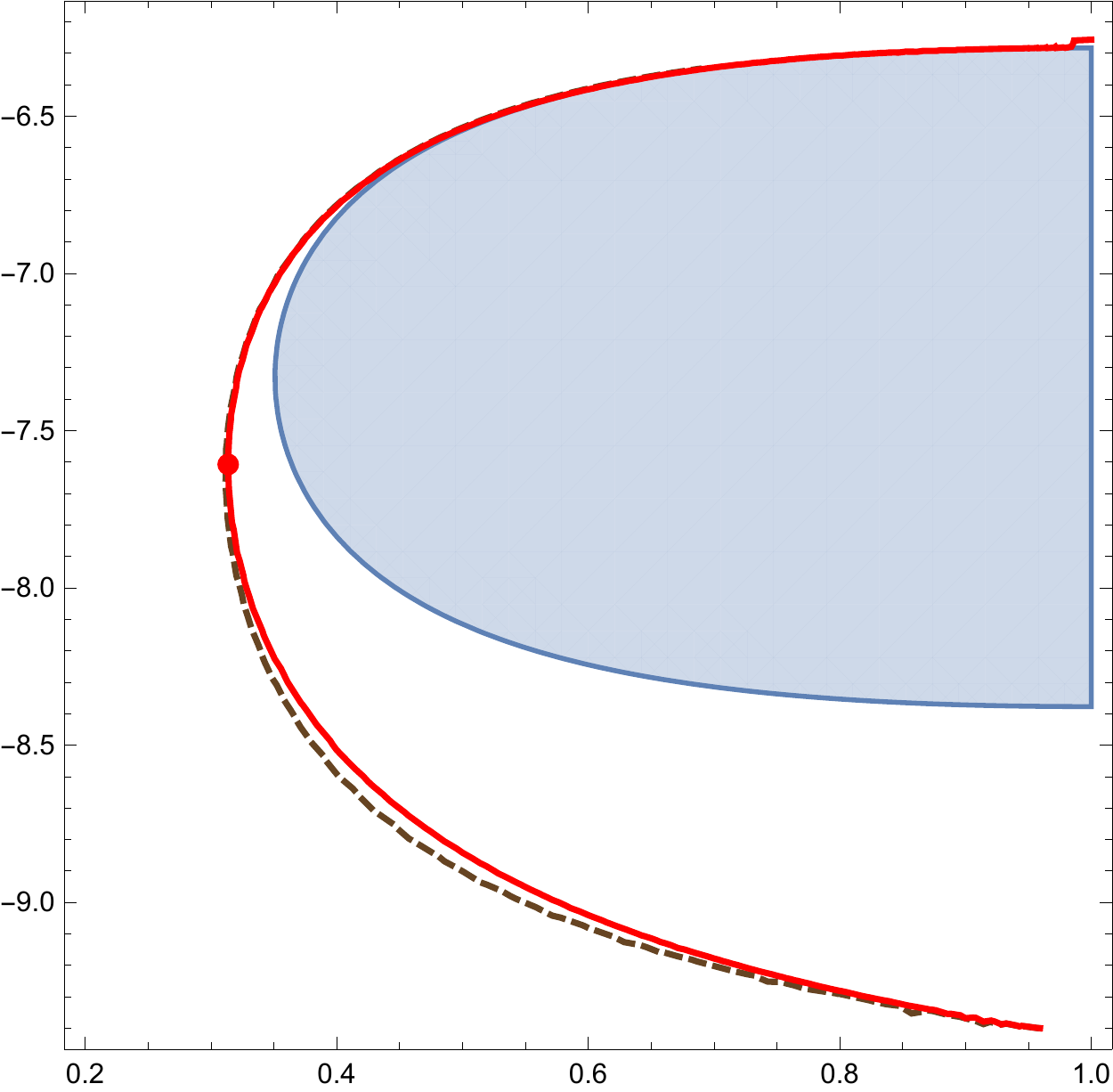}
	\caption{The blue region cover a sector of the $(y,\psi)$-plane where the Bethe approach of \cite{Benini:2018mlo, Benini:2018ywd} predicts exponentially growing behavior. The dashed curve is the deconfinement curve obtained in our approach. The red line is the gravitational curve $\Re F_{\textrm{BH}}=0$. }
	\label{comparisonBM}
\end{figure}

The minimal value of $y$ on the curve is
\be (y, \psi)^{(2)}\approx(0.312,-(2 \pi+1.355))\ed\ee
We can compare this numerical value with the Hawking--Page transition point \eqref{php} which is shown in red in figure \ref{curvecomparison1}.
The computation to order $a_2^2$ leads to a  deconfinement curve remarkably close to the one predicted by the Hawking-Page transition. 

Finally, we would like to compare our results with the large-$N$ analysis of the $\cN=4$ superconformal index which has been recently described in \cite{Benini:2018mlo, Benini:2018ywd} using a Bethe-Ansatz inspired formula. In this language, the dominating saddles can be thought of as solutions of an elliptic Bethe-type equation. The authors of \cite{Benini:2018mlo, Benini:2018ywd} have been able to estimate analytically in which region of the complex $(y,\psi)$-plane the index is expected to have exponential growth, we show such region in figure \ref{comparisonBM}. One virtue of this approach is that, in the region of confidence, the Bethe-Ansatz analysis allows to compute exactly the coefficient of the $\cO(N^2)$ term. The region where the Bethe approach predicts exponentially growing behavior is included in the region where our analysis shows that deconfinement has taken place. 

\newpage

\bigskip
\begin{center}
	\large\textbf{Acknowledgments}
\end{center}
\vspace{-3pt}

\noindent We are grateful to D. Cassani, A. Dymarsky, O. Lisovyy, A. Nedelin, S. Razamat, and T. Sulejmanpasic for discussions. C.C. gratefully acknowledges support from the Simons Center for Geometry and Physics at which some of the research for this paper was performed. C.C. is supported by FPA2015-65480-P and by the Spanish Research  Agency through  the  grant  IFT  Centro de Excelencia Severo Ochoa SEV-2016-059 and a ``La Caixa-Severo Ochoa” international predoctoral grant. A.G., Z.K., and L.T. are supported in part by the Simons Foundation grant 488657 (Simons Collaboration on the Non-Perturbative Bootstrap) and the BSF grant no. 2018204. A.G. is also partially supported by SNSF, Grant No. 185723.  Any opinions, findings, and conclusions or recommendations expressed in this material are those of the authors and do not necessarily reflect the views of the funding agencies.

\bigskip

\appendix

\section{Some Details on Matrix Models}\label{App1}
Throughout the main body of this work we have discussed planar solutions of (holomorphic) matrix models. In this appendix we collect some facts about the computations performed and the general formalism.

Consider a matrix model of the form
\begin{equation} \label{Zgen}
	Z[g,N] = \int_{\Gamma}\rd^N  z_i   \prod_{i<j}\ \Delta^2(z_i-z_j) \exp \left(- N  \sum_{i=1}^N \left(g V(z_i)-\log z_i\right)\right), \,\quad  {\Re g >0}\ec
\end{equation} 
where $\Delta$ is the Vandermonde determinant\footnote{From this point of view, the logarithmic term in the potential is needed to get the right Vandermonde determinant on the unit circle.}, $V(z)$ is a single trace potential and  $\Gamma$ is a contour in the complex plane. We assume that for real, positive coupling $g$,
$\Gamma$ can be deformed onto a negatively-oriented unit circle in the complex plane so that so that \eqref{Zgen} defines an ``analytic continuation" of the unitary matrix integral, see for instance \cite{Alvarez:2016rmo,medinaalvarez2013scurve, Lazaroiu:2003vh, Felder:2004uy} for a nice exposition.
In the case of the GWW model we have
\be V(z)=z+{1\over z}\ec\ee
where the variable $z$ is related to $\alpha$ in \eqref{GWWalpha} by \be z_i=\re^{\ri \alpha_i}\ed\ee
 The large $N$  saddles of the model have the eigenvalues lying on a (possibly disconnected) curve $\gamma$ in the complex plane. We call the number of connected components ``number of cuts''. In the continuum limit this is described by a positive, normalized density $\rho(z)$ supported on $\gamma$.
 More precisely $\rho(z)$ is defined by 
 \be \frac{1}{2 \pi} \text{Disc}_{\gamma} y(z) dz = \rho(z) |dz|\ec\ee
 where $|dz|$ is the line element on $\gamma$ and $y(z)$ is  the spectral curve namely
 \begin{equation}
	y^2(z) = \left(V'(z) \right)^2 - 4 \int_{\gamma} \frac{V'(z)-V'(z')}{z-z'} \rho(z) |dz| \, . \label{eom app}
\end{equation}

\begin{figure}[t]
	\centering
	\includegraphics[width=1\linewidth]{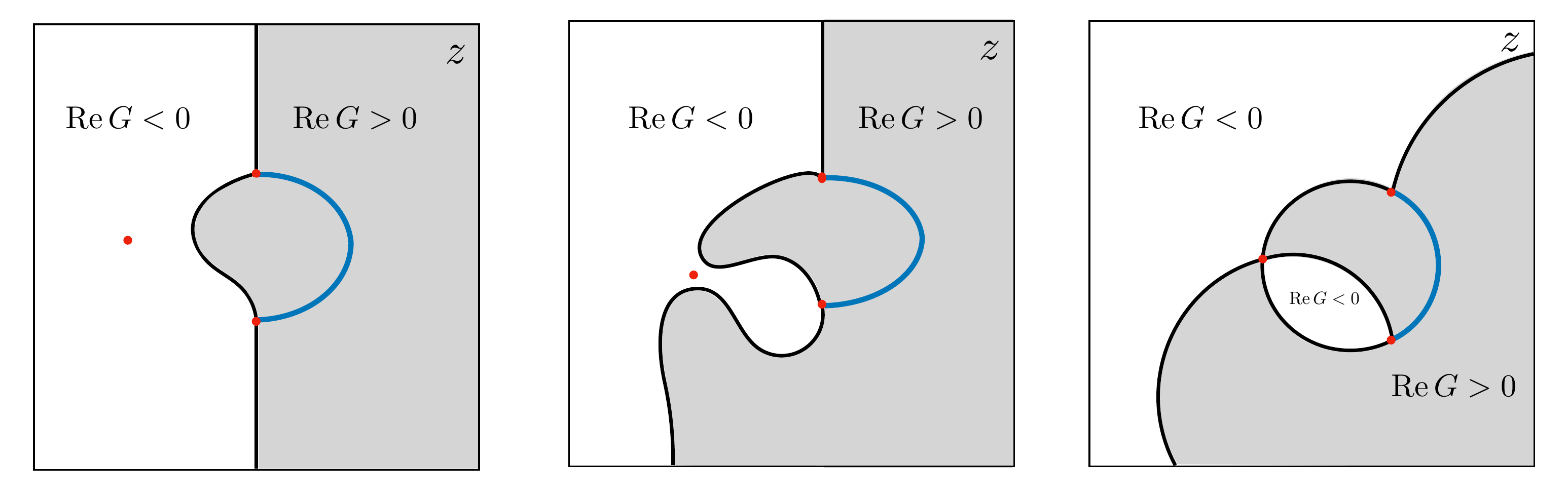}
	\caption{Each image  corresponds to a different value of $g$. In the leftmost figure  $\Re(A^s(g))\gg 0$ while in the rightmost figure $\Re(A^s(g))=0$. The black and blue lines represent the region $\Re G(z)=0$. The blue line denotes the position of the cut, while red dots are the zeroes of $y(z)$ in \eqref{oneap}.
Notice that a ``pinching'' of two lines occurs precisely at the instanton condensation point.
}
	\label{Gfunction}
\end{figure}

The curve  $y(z)$ has square root singularities which signal the beginning of a cut, we call these  ``endpoints.'' For real, positive coupling the cuts belong to the real line and go from one endpoint to another. Likewise, for complex couplings the shape of the cuts is fixed  by demanding that $\rho$ is real and positive along the cuts.
A nice way to put this into equations is to define a function (see \cite{David:1990sk})
\begin{equation}
	G^a(z) = \int_a^z y(z) dz \, ,
\end{equation}
where $a$ is one of the endpoints. Then 
\begin{equation}\label{cut}
	\Re(G^a(z))=0\ec \qquad   z\in \gamma\ed 
\end{equation}
Furthermore, if $\gamma$ defines a branch-cut of the spectral curve, $\Re(G^a(z))$ has the same sign slightly above and slightly below $\gamma$.\\
As an example, let us consider the one-cut phase of the GWW model.
We have
\begin{equation} \label{oneap}
y(z) dz = \frac{g}{2}\sqrt{(z-a)(z-b)} \ \frac{z+1}{z^2} dz \ec
\end{equation}
with end-point solving $a b = 1$ , $a + b =2( 1- 2/g )$. We can find the function $G(z)$ by explicit integration
\begin{equation}\label{Ggww}
x= \frac{g}{2}\frac{z-1}{z} s(z) + \frac{1}{2} \log \left[ \frac{s(z)-(1-A z)}{s(z)+(1-A z)} \ \frac{s(z) +(z-A)}{s(z)-(z-A)}\right] \, ,
\end{equation}
where
$A= (a + b)/2$ and $s(z) = \sqrt{(z-a)(z-b)}= \sqrt{z^2 +1 -2 A z}$. 
Suppose we take $g=|g|\re^{\ri \phi} \in \IC$, $\Re(g)>0$ and study the shape of the cut. One can plug \eqref{Ggww} into \eqref{cut} and solve the latter numerically. The result is shown in figure \ref{Gfunction}.
It is interesting to note that at $z=-1$ the one-cut solution \eqref{oneap} pinches and one has  \be { G(-1)}=A^s(g) \ee
where $A^s$ is the instanton action \eqref{Asdef}, as expected from \cite{Seiberg:2003nm,Kazakov:2004du}. This means that \eqref{cut} evaluated at $z= -1$ corresponds to a line  where instanton events condense.
This singularity can be resolved either by continuing the one-cut solution or by splitting the point $z=-1$ creating a second cut as shown in Fig.~\ref{resolution}. 
If $g$ is such that the instanton action becomes negative, the latter possibility is energetically favorable.

\begin{figure}[t]
	\centering
	\includegraphics[width=0.8\linewidth]{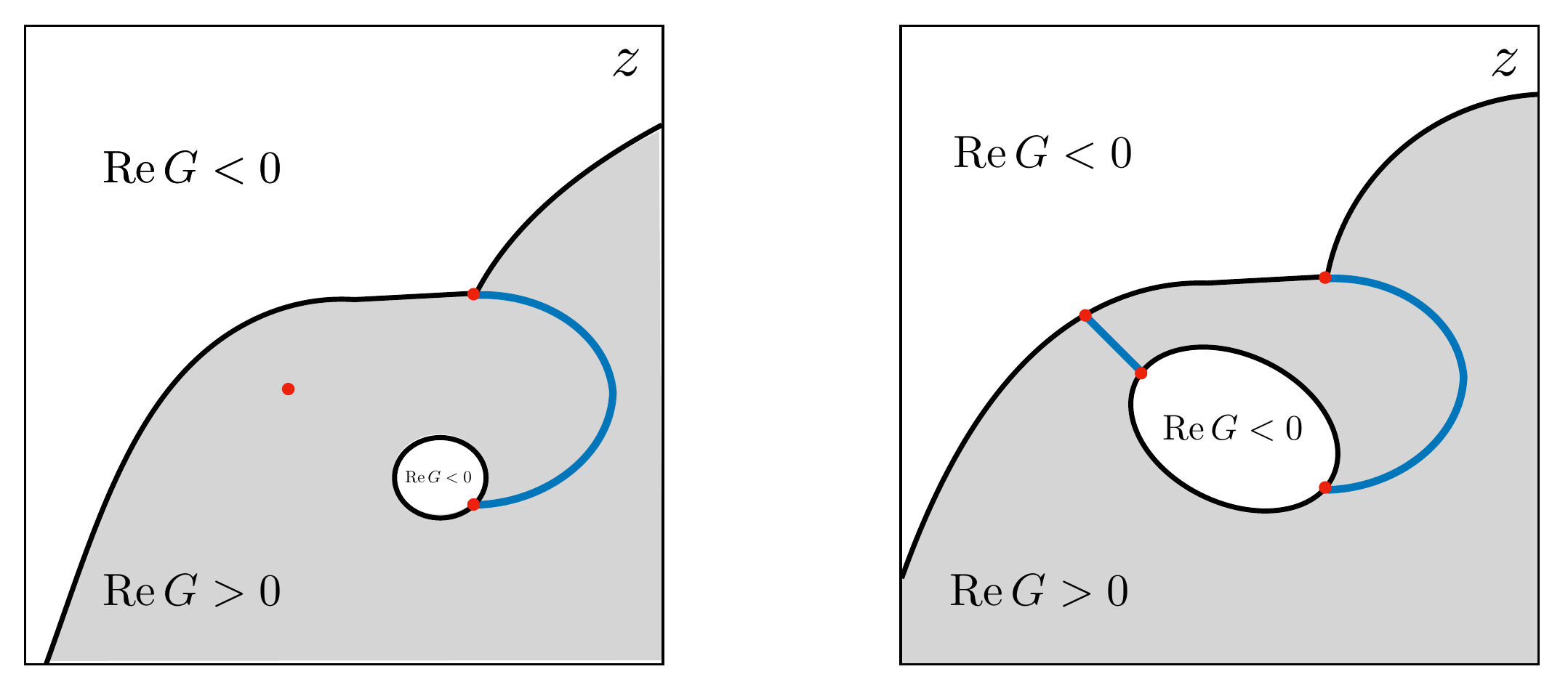}
	\caption{Possible resolutions of the ``pinching'' singularity. On the left, the one-cut solution is continued past it. On the right the $z=-1$ zero of $y(z)$ splits and a new cut opens. Here we consider a value of $g$  where the instanton action is negative, hence the second process is favorable. }
	\label{resolution}
\end{figure}

Alternatively one can construct an explicit parametrization for the cut. Let us take $g=|g|\re^{\ri \phi} \in \IC $, $\Re(g)>0$. We want to find a suitable path \be z(t)\ec \qquad t \in (t_i, t_f) \subset \mathbb{R}\ed \ee 
such that the one-cut density
\be\label{densita} \rho(t) \rd t =\frac{g}{2 \pi \ri}\frac{\left(z(t)^{1/2}+z(t)^{-1/2}\right)}{z(t)}  \sqrt{z(t)+\frac{1}{z(t)}+\frac{4}{g}-2} \ {\rd z(t)\over \rd t}\rd t \ec\ee
is positive and  normalized to 1. Moreover we also demand that \be \label{end}z(t_i) z(t_f)=1\ec \qquad z(t_i)+{z(t_f)}=2-{4\over g}\ed \ee
As an Ansatz we take
\be \label{change} z(t)=\left(\sqrt{1-t^2 e^{-i \phi }}+i t e^{-\frac{1}{2} (i \phi )}\right)^2\ec \qquad t\in (-{1\over \sqrt{|g|}}, {1\over \sqrt{|g|}})\ed\ee
By using \eqref{change} is  it is easy to see that
\be   \rho(t) \rd t=\frac{2 |g| \sqrt{\frac{1}{|g|}-t^2}}{\pi }\rd t\ec\ee
is clearly positive and that equation \eqref{end} is satisfied.

Let us discuss such change of variable more carefully:
\begin{itemize}
\item The function  $z(y)$ in \eqref{change} has branch cuts  for $\phi=0$ at $y=(-\infty, -1)$ and  $y=(1, \infty)$.  This excludes the real segment $ g \in [0,1]$ where the gapless solution exists and we restrict our Ansatz accordingly.  
\item The density \eqref{densita} vanishes by construction at $t_i$ and $t_f$ but also  for $t_s$ such that $z(t_s)=-1$. Hence we have to make sure $t_s$ does not belong to our path. It is easy to show that
\be z(t_s)=-1 \qquad \rightarrow \qquad t_s=\pm \re^{- \ri \phi/2}.  \ee  Hence the point $t_s$ belong to the path only for $ g= 1$ which we have to exclude.
 \end{itemize}
Again, this shows that for $g \in [0,1]$ the path hits the singular point at $t_s$
We can also check that
  \be {\rm Re}(G(z(t)))=0\ec \qquad t \in (t_i, t_f)\ec\ee
and that, on the two sides of the path, $\Re(G(z(t)))$ has the same sign.  
   
\section{Computing the Large-\texorpdfstring{$N$}{N} Free Energy}\label{apfree}

In this appendix we review the computation of the genus zero free energy 
\begin{equation}
F_0= \lim_{N\to \infty}\log N^{-2} Z  \, ,
\end{equation}
fora generic potential of
the form \be V(U) = -\sum_{k=1}^n \frac{g_k}{2} \left(\Tr U^k + \Tr U^{-k}\right).\ee
These were also studied in \cite{Jurkiewicz:1982iz,Mandal:1989ry}. 
It is convenient to define the expectation values of the holonomies:
\begin{equation}
\Omega^k \equiv N^{-1} Z^{-1} \int [D U] \Tr U^k  \exp \left(- N \left(\Tr V(U)+ \Tr \log (U)\right)\right) = \partial_{g_k} F_0 \, .
\end{equation}
or,
\begin{equation}
\Omega^k = \frac{1}{4\pi i} \int_{\mathcal{C}_{\gamma}} z^k y(z) dz \, ,
\end{equation}
where $\mathcal{C}_{\gamma}$ is  a closed contour encircling the cut $\gamma$. 
Hence  the free energy can be derived easily from $y(z)$ by Taylor expansion plus an integration over 't Hooft couplings.

In the following we show how to derive the saddles which are in the body of the paper.

\subsection{Ungapped Saddles}
These saddles have only a single cut which is a closed curve. They are characterized by an Ansatz:
\begin{equation}
y^2_{0-cut}(z) = \frac{H^2(z)}{z^{2 n+2}}
\end{equation} 
with $H(z)$ a degree $2n$ polynomial. In this case the necessity for the cut comes from the need to choose a different sign of the square root between zero and infinity. The equations of motion are easily solved (say near $z=0$) to give
\begin{equation}
y(z) dz = \frac{dz}{z} \left(1 + \frac{g_1}{2} (z+ z^{-1}) + g_2 (z^2 + z^{-2}) + ... + n \frac{g_n}{2} (z^n + z^{-n})\right) \, , \label{AppUngapped}
\end{equation}
for real couplings this gives a density on the circle
\begin{equation}
\rho(\theta)= \frac{1}{2 \pi} \left(1 + 2 \sum_{k=1}^n k \frac{g_k}{2} \cos(k \theta)\right) \, ,
\end{equation} 
and
\begin{equation}
\Omega^k = k \frac{g_k}{2} \, ,
\end{equation}
which gives a free energy
\begin{equation}
F = \sum_{k=1}^n k \frac{g_k^2}{4} \, ,
\end{equation}
having fixed the integration constant to zero by normalizing to the model with a trivial potential.

\subsection{Gapped Saddles}
These saddles will have a single open cut in the complex plane. We start from the Ansatz (symmetric under $z \to 1/z$):
\begin{equation}
y^2_{1-cut}(z)= \frac{1}{z^{2n +2}} \left(z^2 +1-A z\right)\left[ \alpha_0 (z^{2n-1}+1) + \alpha_1 (z^{2n-2}+z) + ... + \alpha_{n-1}(z^{n}+z^{n-1})  \right]^2 \, ,
\end{equation}
The cut endpoints are located at $a=1/b = \frac{A}{2} - \sqrt{\left(\frac{A}{2}\right)^2-1}$ and te right asymptotics to fulfill the equations of motion.
The equations of motion expanded near zero up to order $1/z$ give:
\begin{equation}
\sum_{k=0}^r C_{r-k}^{-1/2}(A/2) \alpha_k = \delta_{r,0} \frac{n}{2}g_n + \delta_{r,1} \frac{(n-1)}{2} g_{n-1} + ... + \delta_{r,n} \, , \ \ \ r=0,...,n \, , \alpha_{n}=\alpha_{n-1} \, ,
\end{equation}
with $C_{r-k}^{-1/2}(A/2)$ the Gegenbauer polynomials. These can easily be solved recursively case y case or recasted as a matrix equation by defining $M_{i,j}= C^{-1/2}_{i-j}(A/2)$, with lower-triangular $M$. The final equations for the endpoints are given by a degree $n$ polynomial in $A$, which cannot be solved analytically in full generality. We give some examples:
\paragraph{\underline{n=1}}
This is the GWW model:
\begin{equation}
\alpha_0 = \frac{g}{2} \, , \ \ A= 2\left(1- \frac{2}{g}\right) \, ,
\end{equation}
Expanding $y(z)$:
\begin{equation}
\Omega= 1 - \frac{1}{2 g} \, .
\end{equation}
Matching with the zero-cut solution at the point at $g=1$ (where the cut closes):
\begin{equation}
F =g - \frac{1}{2} \log(g) - \frac{3}{4} \, .
\end{equation}
\paragraph{\underline{n=2}}
This is the model of \ref{twoterms}, for ease of confrontation we set $g_2=h$.
We find:
\begin{equation}
\alpha_0 = h \, , \ \alpha_1 = h \frac{A}{2} +\frac{g}{2} \, ,
\end{equation}	
and the endpoint equation
\begin{equation}
1-\frac{h}{2} +\frac{A^2 h}{8} -\alpha_1 + \frac{A \alpha_1}{2}=0
\end{equation}	
these have two solutions
\begin{align}
A &= \frac{1}{3} \left(2 -\frac{g}{h} \pm \xi    \right)= A_{\pm} \, , \\
\alpha_1 &= \frac{h}{3}\left(1 + \frac{g}{h} \pm \frac{\xi}{2}\right)  = \alpha_{\pm}
\end{align}
Being $\xi = \sqrt{(g+4 h)^2 -24 h}$.
Looking for a solution which reduces to the $n=1$ case as $h\to 0$ we see that we need to choose the $+$ branch.

To compute the free energy we expand $y(z)$ to find:
\begin{align}
\Omega^2&= \frac{-g^4-8 g^3 h+36 g^2 h+128 g h^3+(g-4 h) \xi^3+256 h^4+288 h^3-216 h^2}{864 h^3} \, , \\
\Omega &= \frac{g^3+12 g^2 h+48 g h^2-\xi^3-36 g h+64 h^3+288 h^2}{432 h^2}
\end{align}
integrating over $h$ and using the $n=1$ result to fix the remaining ambiguity gives \eqref{f0gb}.

\subsection{Two-Cut Saddles in GWW}\label{2cutApp}
Two cut saddles have a much richer (and complicated) structure. Apart from symmetric choices we have no closed form for their free energy. We can however give a series expansion at both strong and weak coupling. This is possible thanks to an underlying connection between the GWW model and  Seiberg-Witten theory. Indeed form \cite{Eguchi:2010rf} it follows that  the GWW spectral curve coincides with that of the $SU(2)$ theory with $N_f=2$ massive hypers upon matching parameters.

Denoting by $n_s \equiv 1- {N_+\over N}$, where $N-N_+$ is the number of eigenvalues in the second cut, at small $g$ we have:
\be  \label{fow} 2 F_{0}^{(e)}(g, n_s)= \frac{g^2 \left(n_s^2-2 n_s+2\right)}{4 (n_s-1)^2}+n_s (n_s \log (g)-2 \log (g)-2+\log (4))+n_s\sum_{n\geq 2} g^{2n}c_n(n_s)\ee
where the first two $c_n(n_s)$ coefficients are 
\be \ba 
c_2(n_s)&=\frac{(n_s-4) n_s^2+8}{128 (n_s-1)^6}\ec\\
c_3(n_s)&=\frac{(n_s-2) (5 (n_s-2) n_s-4)}{384 (n_s-1)^{10}}\ed\ea\ee
The expression \eqref{fow}  corresponds to electric expansion in SW theory as in \cite[eq (3.19)]{Eguchi:2010rf}.\footnote{In \cite[eq (3.19)]{Eguchi:2010rf} one has the freedom to fix an overall $g$-independent integration constant. Here we fix it to be $n_s (2-\log (4))$ so that it agrees with \eqref{awf}} The dictionary between   \cite[eq (3.19)]{Eguchi:2010rf} and our expression \eqref{fow} is given by 
\be\ba
\Lambda \to & \quad  g , \\
{a\over m} \to  & \quad 1-n_s , \\
m \to & \quad 1 .
\ea \ee
When $n_s=0$ we recover the known  ungapped one-cut solution
\be F_0 (g, 0)= \left( {g^2\over 4}\right).\ee
Moreover we find that
\be \label{awf}A^{\textrm{w}}(g)= 2 \partial_{n_s}F_0^{(e)}(n_s,g) \mid_{n_s=0} \ee
where $ A^{\textrm{w}}$ is given in \eqref{Aw}.
We have checked \eqref{awf} in the small $g$ expansion up to $\mathcal{O}(g^{16})$. This kind of relations are expected from a matrix model perspective, see for instance \cite{Marino:2007te} and reference therein. However, to our knowledge, this was never worked out in the literature for the specific example of the GWW model.

The expansion of the free energy at  large $g$ is then mapped to the  magnetic expansion in Seiberg-Witten theory. We get
\begin{equation}\label{magnetic}
\begin{aligned}
F_{0}^{(m)}(g, n_s) &= g (2 n_s+1) +\frac{1}{4} (-2 n_s (n_s+1) (3+\log (16))-3)  \, \\
&+\frac{1}{4} \left(-2 n_s^2 \log \left(\frac{g}{n_s}\right)-2 (n_s+1)^2 \log \left(\frac{g}{n_s+1}\right)\right) \\
&+ n_s \sum_{n=1}^{\infty} \frac{1}{g^{n}} \tilde{\varphi}_n(n_s) \, ,
\end{aligned}
\end{equation}
where the first two $\tilde{\varphi}_n(n_s)$ coefficients are 
\be \label{phi} 
\ba
\tilde{\varphi}_1(n_s)&= \frac{1}{4} (n_s+1) (2 n_s+1) \, , \\
\tilde{\varphi}_2(n_s) &= \frac{1}{16} (n_s+1) (5 n_s (n_s+1)+1) \, . 
\ea 
\ee
These are easily computed by using \cite[eq. (A.33)]{Bonelli:2016qwg}\footnote{
A similar connection has appeared in \cite{Dunne:2019aqp} but without reference  to the GWW two-cut phase.}.
The dictionary between \cite[eq. (A.33)]{Bonelli:2016qwg}  and the expression \eqref{phi} is given by 
\be\ba
s \to &  -4 \ri g , \quad
\theta_2=-\theta_1 \to &  1 ,\quad
\nu \to & \quad 2 n_s+1.
\ea \ee
When $n_s=0$  we recover the gapped one-cut phase:
\be{  F_{0}(g,0) =  g-{3/4}-{1\over 2}\log g .}\ee 
Moreover we find that
\be \label{aws}A^{\textrm{s}}(g)= \partial_{n_s}F_0^{(m)}(n_s,g) \mid_{n_s=0} \ee
where $ A^{\textrm{s}}$ is given in \eqref{Asdef}.
We have checked \eqref{aws} in the large $g$ expansion. 
To our knowledge, this kind of relations have never been worked out in the literature for the specific example of the GWW model.

\newpage
\bigskip
\renewcommand\refname{\bfseries\large\centering References\\ \vspace{-0.4cm}
\addcontentsline{toc}{section}{References}}
\bibliographystyle{utphys.bst}
\bibliography{BlackHoleAdS_Final.bib}
	
\end{document}